\documentclass[usenatbib]{mn2e}
\usepackage{graphicx}
\usepackage{amsbsy}
\usepackage{xspace}
\usepackage{subfigure}
\usepackage[toc,page]{appendix}
\usepackage[colorinlistoftodos]{todonotes}
\usepackage{hyperref}
\usepackage{booktabs,caption}
\usepackage[flushleft]{threeparttable}
\usepackage{soul}

\def\gtrsim{\mathrel{\hbox{\rlap{\hbox{\lower5pt\hbox{$\sim$}}}\hbox{$>$}}}}

  \title[Luminous IR Variables]{SPIRITS Catalog of Infrared Variables: Identification of Extremely Luminous Long Period Variables}
  \author[SPIRITS Collaboration]
  {\parbox{18cm}{V.~R.~Karambelkar$^{* 1,2}$, S.~M.~Adams$^{1}$, P.~A.~Whitelock$^{3,4}$, M.~M.~Kasliwal$^{1}$, J.~E.~Jencson$^{1}$, M.~L.~Boyer$^{5}$, S.~R.~Goldman$^{5}$, F.~Masci$^{6}$, A.~M.~Cody$^{7}$, J.~Bally$^{8}$, H.~E.~Bond$^{5,9}$, R.~D.~Gehrz$^{10}$, M.~Parthasarathy$^{11}$, \& R.~M.~Lau$^{12}$ }
  \\
  \\
  $^{1}$ Cahill Center for Astrophysics, California Institute of Technology, Pasadena, CA 91125, USA\\
  $^{2}$ Department of Physics, Indian Institute of Technology Bombay, Mumbai 400076, India\\
  $^{3}$ South African Astronomical Observatory, PO Box 9, 7935 Observatory, South Africa\\
  $^{4}$ Department of Astronomy, University of Cape Town, Private Bag X3, Rondebosch 7701, South Africa\\
  $^{5}$ STScI, 3700 San Martin Drive, Baltimore, MD 21218, USA\\
  $^{6}$ Infrared Processing and Analysis Center, California Institute of Technology, Pasadena, CA 91125, USA\\
  $^{7}$ NASA Ames Research Center, Moffett Field, CA 94035, USA \\
  $^{8}$ Center for Astrophysics and Space Astronomy, University of Colorado, UCB 389 Boulder, CO 80309, USA \\
  $^{9}$ Department of Astronomy and Astrophysics, Pennsylvania State University, University Park, PA 16802, USA \\
  $^{10}$ Minnesota Institute for Astrophysics, 116 Church Street S. E., Minneaopolis, MN 55455, USA\\
  $^{11}$ Indian Institute of Astrophysics, Koramangala, Bangalore 560034, India\\
  $^{12}$ Institute of Space \& Astronautical Science, Japan Aerospace Exploration Agency, Sagamihara, Kanagawa 252-5210, Japan\\
  * E-mail: viraj.k@iitb.ac.in
 }
\begin{document}
\voffset -1.5cm
\maketitle

\begin{abstract}
We present a catalog of 417 luminous infrared variable stars with periods exceeding 250 days. These were identified in 20 nearby galaxies by the ongoing SPIRITS survey with the \emph{Spitzer Space Telescope}. Of these, 359 variables have $M_{[4.5]}$ (phase-weighted mean magnitudes) fainter than $-12$  and periods and luminosities consistent with previously reported variables in the Large Magellanic Cloud. However, 58 variables are more luminous than $M_{[4.5]} = -12$, including 11 that are brighter than $M_{[4.5]} = -13$  with the brightest having $M_{[4.5]} = -15.51$. Most of these bright variable sources have quasi-periods longer than 1000 days, including four over 2000 days. We suggest that the fundamental period-luminosity relationship, previously measured for the Large Magellanic Cloud, extends to much higher luminosities and longer periods in this large galaxy sample.
We posit that these variables include massive AGB stars (possibly super-AGB stars), red supergiants experiencing exceptionally high mass-loss rates, and interacting binaries. We also present 3.6, 4.5, 5.8 and 8.0 $\mu$m photometric catalogs for 
all sources in these 20 galaxies.
\\
\\
\end{abstract}


\section{Introduction}
The parameter space of stellar optical variability has been explored for centuries, revealing many classes of variable stars \citep[see, e.g.,][for a review]{GaiaHRvariables}.  However, significant ($>$0.3 mag) variability with long periods ($>$200 d) in luminous (brighter than $M_{\mathrm{bol}}=-4$) stars is primarily limited to Luminous Blue Variables (LBVs), pulsating Red Supergiants (RSGs), and Asymptotic Giant Branch (AGB) stars \citep{Samus2017}.
These three classes of Long-Period Variables (LPVs), which have periods longer than a few hundreds of days, can be prolific dust producers and require observations in the infrared (IR) to understand their spectral energy distributions (SEDs) and dust production, but until recently there have been few IR variability surveys with sufficiently long temporal baselines.


 

AGB stars constitute a large proportion of long period IR variables and the largest amplitudes are found among the Mira variables, which tend to be at the tip of the AGB \citep{Soszynski2009,Whitelock17}. These variables have been studied by various surveys of the Local Group. Working in the near-IR ($JHK(L)$), \citet{Feast1989} and \citet{Hughes1990} discovered and studied numerous LPVs in the Large Magellanic Cloud (LMC). This has gradually been extended to other Local Group galaxies \citep[e.g.][and references therein]{Menzies2015, Whitelock2018}.  The Optical Gravitational Lensing Experiment \citep[OGLE;][]{Udalski1993} has also characterized large numbers of LPVs in the Magellanic Clouds and elsewhere \citep[e.g.][]{Soszynski2009}, with sufficient data to derive excellent periods. Recently, \citet{Ita2018} presented $JHK$ time series data for the central region of the Small Magellanic Cloud (SMC).  

Using 8 bands of IR photometric data from the Surveying the Agents of Galaxy Evolution \citep[SAGE;][]{Meixner06} survey and periods from the Massive Compact Halo Object \citep[MACHO;][]{Alcock1993} survey, \citet{Riebel10} reported the IR period-luminosity (PL) relations for around 30,000 AGB stars in the LMC. \citet{Riebel15} employed the variability criteria described in \citet{Vijh09} to identify variables in the LMC and SMC using 3.6 and 4.5 $\mu$m data. They found 10 new dust-obscured large-amplitude AGBs in the LMC and six in the SMC. These objects are not detected by optical surveys, and hence do not have a measured optical variability. Dust in Nearby Galaxies with \emph{Spitzer} \citep[DUSTiNGS;][]{Boyer15} is a 3.6 and 4.5 $\mu$m survey of 50 dwarf galaxies to identify dust producing AGB stars. Using two epochs and the variability criteria defined by \citet{Vijh09}, this survey identified 710 variables in these galaxies. 

AGB stars can be classified into carbon-rich and oxygen-rich variables based on their surface C/O-ratios.  This can be done using spectroscopy or narrow-band photometry.  AGB stars with the reddest $[3.6]-[4.5]$ colors are considered ``extreme" AGB stars (eAGB; \citealp{Thompson09,Boyer15}). While this is a normal phase of stellar evolution, eAGBs are characterized by the extreme amounts of dust enshrouding them, and are a major contributor to the dust content of the ISM \citep{Matsuura2009,Riebel12}.

AGB stars can also be classified based on their periods. The periods of large-amplitude AGB variables are typically longer than 100 days. These stars can oscillate in their fundamental modes (as do Miras)
or exhibit overtones. The fundamental modes have periods ranging from 100 - 1000 days, while the overtones are characterized by shorter periods. AGB variables exhibit a linear log(period) $-$ log(luminosity) relationship. The different modes of oscillation can be separated into six sequences in the period luminosity space \citep{Wood1999}. \citet{Menzies2019} recently summarized what is known about Miras with periods longer than 1000 days. There are only 17 known in the Galaxy, 18 in the LMC and three in the SMC. The longest period (1859 days) is for MSX-SMC-055, which is discussed in \S 3.

Super-AGB stars are a subclass of AGB stars at the upper end of the AGB mass range (8-12~$M_\odot$). These stars are expected to evolve similarly to the slightly less massive oxygen-rich AGB stars, but ultimately explode as electron-capture supernovae or produce ONe white dwarfs. Super-AGB stars are also expected to pulsate, populating the high-luminosity end of the AGB period-luminosity relationship for fundamental mode pulsators. While super-AGB stars have a strong theoretical backing \citep{Siess2007,Doherty+2015, Doherty+2017}, it has been difficult to identify them observationally. Their observed properties tend to be very similar to red supergiants and other massive AGB stars, and they are also expected to be quite rare. The IR pulsation properties provide an additional clue towards confidently identifying these objects. The amplitudes of variability of these super-AGB stars are expected to be larger than those of RSGs, and may be a key for distinguishing between the two. We discuss the AGB stars in our sample in $\S$\ref{sec:agb_candidates} and the potential super-AGBs in section $\S$\ref{sec:brightest_sources}.

Dusty oxygen-rich AGB and RSG stars in their final evolutionary stages exhibit circumstellar OH maser emission. These stars are obscured in the optical, but bright in the IR due to the large quantity of dust surrounding them. These stars are designated as OH/IR stars. The OH maser emission is strongest at 1612 MHz. \citet{Goldman17} and \citet{Goldman18} identify OH/IR stars in the LMC and SMC, and report their periods and amongst other photometric data, their \emph{Spitzer} (\citealp{Werner04,Gehrz07}) - InfraRed Array Camera \citep[IRAC;][]{Fazio04} 3.6 and 4.5 $\mu$m magnitudes.
\citet{Whitelock17} and \citet{Boyer15} recently noted correlations with more luminous stars being dustier (redder $[3.6]-[4.5]$ colors) and dustier stars having larger amplitudes in the IR variability. The evolution of binary stars \citep[e.g.,][]{DeMarco2017} can also give rise to both variability and mass-loss. It seems possible that massive stars in wide binaries, such as $\eta$ Carinae, and common-envelope stars, such as Thorne-$\dot{Z}$ytkow Objects (TZOs) will be among our sources. This is discussed briefly in $\S$\ref{sec:brightest_sources}.

Until recently, the largest IR variability surveys using \emph{Spitzer} were limited to the LMC and SMC 
\citep{Vijh09,Riebel15}, M33 \citep{McQuinn07}, and nearby dwarf galaxies with DUSTiNGS \citep{Boyer15}. These surveys are limited to lower mass galaxies and/or are comprised of a small number of epochs over a small time baseline (requiring the use of variability indices and periods from near-IR/optical surveys).  Accordingly, these surveys do not provide a complete census of the most dusty and obscured IR variables.



Since 2014, the SPitzer InfraRed Intensive Transients Survey \citep[SPIRITS;][]{Kasliwal17} has been monitoring nearly 200 nearby ($d<30$ Mpc) galaxies (typically with cadences of 3--6 months)
using \emph{Spitzer}/IRAC to (primarily) search for IR transients. 
A detailed description of the galaxy sample, depth and cadence of observations of the SPIRITS survey can be found in \citet{Kasliwal17}.  
In this paper we present the IR variable stars identified in the closest and most luminous galaxies in the SPIRITS survey.  In \S\ref{sec:catalogs} we describe our galaxy sample, source catalogs, photometry, period-fitting, and variable classification.  In \S\ref{sec:results} we present our results and suggest that the period-IR luminosity relation measured in the LMC and SMC extends to longer periods and higher luminosities than previously observed.  This relation extends beyond the maximum luminosity expected for AGB stars, suggesting that the most luminous IR variables in our sample may be pulsating super-AGB stars or RSGs experiencing exceptionally high mass-loss rates.


\section{Catalog Construction}
\label{sec:catalogs}
\subsection{Galaxy sample}
\label{sec:gal_sam}
In this paper, we generate point spread function (PSF) catalogs and identify variables in the closest and most luminous galaxies in the SPIRITS survey using a two-tiered selection strategy that includes the 9 galaxies within one Mpc targeted by SPIRITS in addition to the 11 galaxies within 10 Mpc targeted by SPIRITS with B-band absolute magnitudes brighter than $-20.6$. Beyond 10 Mpc, accurate stellar photometry of all but the most luminous stars becomes extremely difficult. Our two-tiered strategy allows for adequate coverage of both : the more common lower luminosity AGBs (that belong to the nearby galaxies) as well as the rare high luminosity variables (that belong predominantly to the distant galaxies). It also enables us to validate our methodology against existing works. Our galaxy sample includes four nearby dwarf galaxies that were targeted by the DUSTiNGS survey.  In \S\ref{sec:dustings} we compare our analysis of variables in the four DUSTiNGS galaxies to that by Goldman et al., in prep. 
Table \ref{tab:galaxy_properties} lists the galaxies in our sample and summarizes their properties.

\begin{table*}
\begin{threeparttable}
\caption{Galaxy Properties}
\label{tab:galaxy_properties}
\begin{tabular}{lcccccccc}
\toprule
 Galaxy &  Number of & \multicolumn{2}{c}{Number of}  & Distance & Number of & Stellar Mass & $B$ mag & SFR\\
&  SPIRITS & \multicolumn{2}{c}{Variables} & (Mpc) & Epochs & ($M_{\sun}$) & (mags) & log($M_{\sun}yr^{-1}$) \\
& sources & \emph{Total} & \emph{Bright$^{~o}$} &  &  &  & &    \\
\midrule
Fornax & 21 & 0& & 0.15 & 12 & 2.0$\times$10$^{7}$ & -11.5 & \\
Leo I & 50 & 1& & 0.26 & 12& 5.5$\times$10$^{6}$ & -11 & \\
NGC~6822 & 200 & 35& & 0.48 & 28 & 1.66$\times$10$^{9}$ & -15.2 & -1.9 \\
NGC~185* & 126 & 56& & 0.63 & 16 & 1.38$\times$10$^{8}$ & -14.7 & -4.73 \\
NGC~147* & 57 & 18& & 0.73 & 13 & 6.2$\times$10$^{7}$ & -14.8 & -6.33 \\
M32$^{\dagger}$ & 171 & 0& & 0.78 & 9 & 3.2$\times$10$^{8}$ & -14.8 & -5.88\\
IC~1613* & 61 & 10& & 0.74 & 28 & 7.6 $\times$10$^{7}$ & -14.5 & -2.35 \\
M110 & 389 & 189& & 0.81 & 14 & 1.91$\times$10$^{8}$ & -16.1 & $<$-7.73 \\
WLM* & 51 & 12& & 0.96 & 24 & 1.86$\times$10$^{8}$ & -14.1 & -2.68 \\
\hline
IC~342 & 216 & 10& 3 & 3.39 & 15 &1.41$\times$10$^{11}$ & -20.7 & -0.26\\
M81 & 64 & 16& 5 & 3.61 & 37& 1.86$\times$10$^{11}$ & -20.9 & 0 \\
M83 & 130 & 52& 33 & 4.66 & 26 &2.09$\times$10$^{11}$ & -20.6 & 0.44\\
NGC~6946 & 29 & 2& 1 & 5.89 & 38 & 7.76$\times$10$^{10}$ & -20.8 & 0.63\\
M101 & 37 & 12& 12 & 6.95 & 33 & 2.24$\times$10$^{11}$ & -21.1 & 0.46\\
M106 & 2 & 1&  1 & 7.31 & 15 & 2.14$\times$10$^{11}$ & -21.2 & 0.44\\
M51 & 22 & 1& 1 & 8.58 & 34 & 4.17$\times$10$^{10}$ & -21.4 & 0.46\\
NGC~6744 & 5 & 2& 2 & 8.95 & 16 &2.24$\times$10$^{11}$ & -21 & 0.35\\
M63 & 13 & 0& & 8.95 & 16 & 2.19$\times$10$^{11}$ & -21 & 0.21 \\
NGC~2903 & 13 & 0& & 9.33 & 15 & 1.35$\times$10$^{11}$ & -20.9 & -2.93 \\
M104 & 8 & 0& & 10.22 & 16 & 5.75$\times$10$^{11}$ & -21.8 & -1 \\
\bottomrule 
\end{tabular}
\begin{tablenotes}
\small
\item The stellar masses, $B$ band magnitudes and integral star formation rates (SFRs)
are from \citet{Karachentsev13}.  The distances are taken from the Cosmicflows-3 catalog \citep{Tully17}, with the exception of NGC~6946, where the distance is from the \citet{Karachentsev13} catalog because it does not have a Cosmicflow-3 distance. \\
* Indicates galaxies that are also included in the DUSTiNGS survey (NGC~185, NGC~147, IC~1613 and WLM).\\$\dagger$ We exclude the variables detected in M32, because the number of epochs for this galaxy is smaller than 10.\\$o$ The bright variables are those with $M_{[4.5]}$ brighter than $-12$ (which is close to the maximum luminosity of the LMC AGB sample examined by \citet{Riebel10}).
\end{tablenotes}
\end{threeparttable}
\end{table*}
\begin{table}
\begin{threeparttable}
\caption{PSF Catalog Properties*}
\label{tab:catalog_properties}
\begin{tabular}{llcccc}
\toprule
Galaxy & [3.6] & [4.5]  & [5.8] & [8.0] & \\
\midrule
Fornax & 48804 & 49868 & 17613 & 16777 &  \\ 
Leo I & 44766 & 43634 & 196$^{\dagger}$  & 63402 &  \\
NGC~6822 & 52262 & 55686 & 101348 & 113344 & \\ 
NGC~185* & 14659 & 14165 & 11625 & 13440 & \\
NGC~147* & 13307 & 12813 & 13629 & 13013 & \\
M32 & 25818 & 28188 & 160818 & 212120 & \\
IC~1613* & 18006 & 21093 & 24176 & 20842 & \\
M110 & 41156  & 40606 & 205870 & 205209 & \\
WLM* & 13169 & 16514 & 19284 & 20465 & \\
\hline
M104 & 35462 & 36667 & 54715 & 62910 & \\
M51 & 33829 & 36326 & 87755 & 93462 & \\ 
M106 & 23874 & 31840 & 28739 & 75601 & \\
M101 & 46770 & 46431 & 124846 & 199673 & \\
NGC~6744 & 16397 & 17089 & 11294 & 20103 &\\
M63 & 41888 & 45237 & 112676 & 121404 & \\
M81 & 44023 & 42785 & 228825 & 242552& \\
NGC~2903 & 58593 & 66706 & 94257 & 95640 & \\
NGC~6946 & 37219 & 38389 & 94773 & 89262 &  \\
IC~342 & 115573 & 123278 & 163581 & 141576 & \\
M83 & 55682 & 57117 & 400578 & 304464 &\\
\bottomrule 
\end{tabular}
\begin{tablenotes}
\small
\item * Number of sources in the catalogs for each channel.
\item $\dagger$ The 5.8$\mu$m reference image for Leo I is shallow resulting in a low number of sources.
\end{tablenotes}
\end{threeparttable}
\end{table}

\subsection{Reference Photometry}
\label{sec:refphot}
Following the approach of \citet{Kasliwal17} we adopt the [3.6], [4.5], [5.8], and [8.0] supermosaic images of each galaxy from the Spitzer Heritage Archive\footnote{\href{http://sha.ipac.caltech.edu/}{http://sha.ipac.caltech.edu/}} as reference images.
We used the DAOPHOT/ ALLSTAR \citep{Stetson87} package to perform PSF photometry on these images. For each reference image, the PSF was constructed using isolated stars in the image. A $2.4\arcsec$ disk was used to measure the brightness of sources and an annulus of $2.4\arcsec - 7.2\arcsec$ was used to estimate sky background while constructing the PSF. We used a PSF fitting radius of 4.2\arcsec, which is roughly 2.5 times the FWHM of the PSF. 

As the PSF was terminated at a finite radius and not allowed to extend to infinity, there is a slight underestimation of the total flux. To correct for this, we followed the procedure described in \citet{Khan15c}. We performed aperture photometry on all sources in our PSF catalog for each image, using IRAF ApPhot/Phot. We used an aperture radius of 2.4$\arcsec$ with a sky annulus of 2.4$\arcsec$-7.2$\arcsec$, and empirically derived aperture corrections of 1.213, 1.234, 1.379 and 1.584 for the 3.6, 4.5, 5.8 and 8.0 $\mu$m channels respectively. The difference between aperture and PSF magnitudes of a source is an indication of crowding of stars around that source. The sources for which this difference is relatively small are relatively isolated in the image. We calculated the mean of the difference between PSF and aperture magnitudes for these relatively isolated sources for each image, and used it as a zeropoint offset to account for the flux underestimation due to finite PSF fitting radius. Following \citet{Khan15c}, we perform this correction procedure for all but the 8.0 $\mu$m images, which are badly affected by crowding.

\begin{figure*}
	\includegraphics[width=\textwidth]{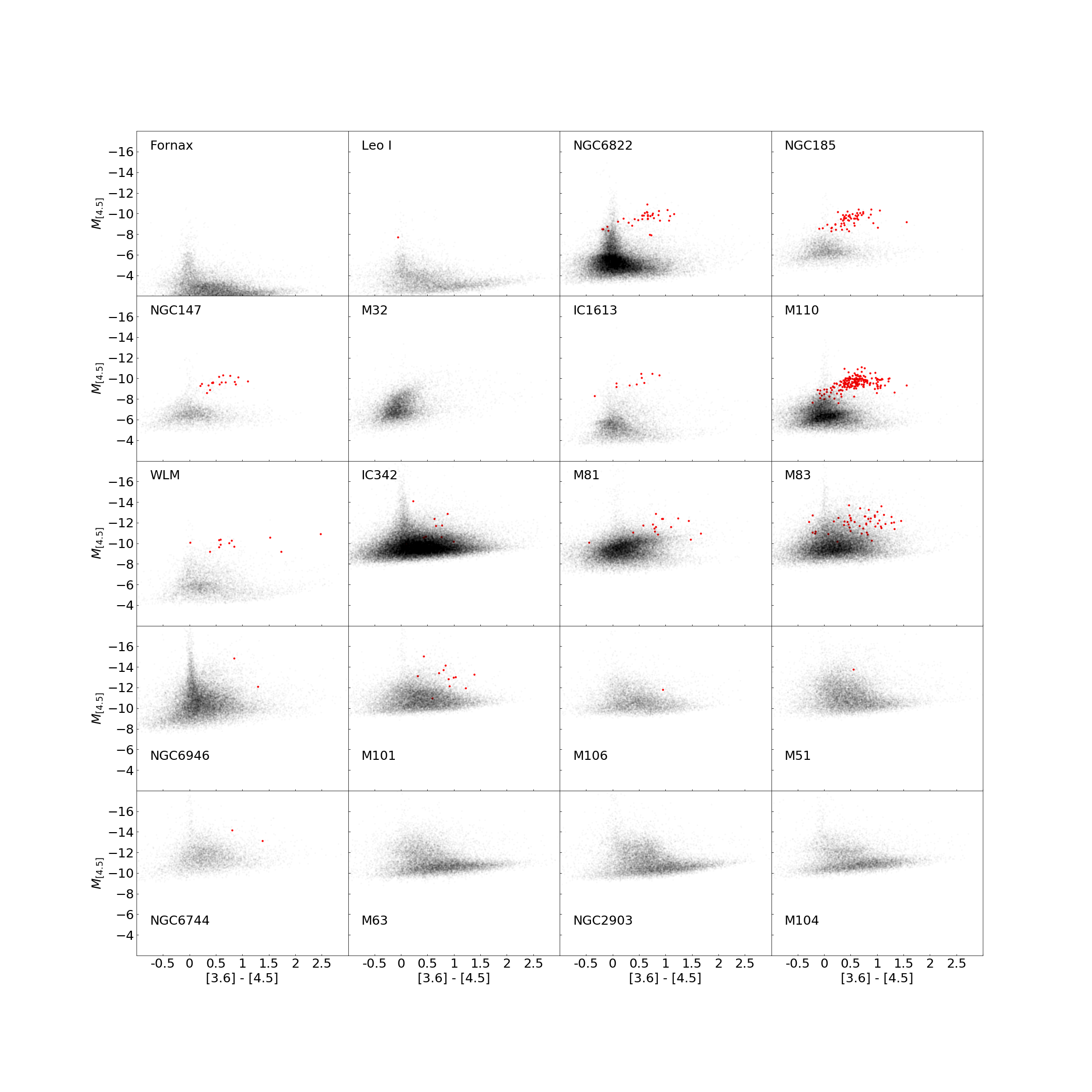}
    \caption{The reference $M_{[4.5]}$ vs $[3.6]-[4.5]$ colors for our variables in each galaxy (red dots in each panel), with the $M_{[4.5]}$ vs $[3.6]-[4.5]$ for the entire galaxy as the background (small black points). To generate the background, we cross-match the 3.6 $\mu$m catalogs with our 4.5 $\mu$m catalogs, using a radius of 1$\arcsec$. We find that most of our variables are among the reddest objects in the galaxies. It is also evident that the reddest variables are generally at the top of the luminosity distribution.}
    \label{fig:all_gal_colmag}
\end{figure*}

We identified all sources that are $>$ 1$\sigma$ brighter than the background and have PSF magnitude uncertainties smaller than 0.3. The instrument magnitudes were converted to apparent magnitudes using the Vega-calibrated zeropoints of 18.80, 18.32, 17.83 and 17.20  for the 3.6, 4.5, 5.8 and 8.0$~\mu$m channels respectively\footnote{\href{http://irsa.ipac.caltech.edu/data/SPITZER/docs/dataanalysistools/}{http://irsa.ipac.caltech.edu/data/SPITZER/docs/dataanalysistools/}}.

The 3.6, 4.5, 5.8 and 8.0 $\mu$m photometric catalogs for all sources in the 20 galaxies are available in Zenodo at \url{https://doi.org/10.5281/zenodo.2643483}. We present the number of sources in each of the catalogs in Table \ref{tab:catalog_properties}.
\citet{Khan15c} and \citet{Khan17} presented PSF catalogs for 22 SPIRITS galaxies (including seven of the galaxies we consider in this paper). We validate our procedure for constructing PSF catalogs by comparing our 3.6 and 4.5 $\mu$m catalogs for the galaxies NGC~2903, M83, NGC~6822, NGC~6946, M51, M101 and M81 to those of \cite{Khan17} using a matching radius of $1 \arcsec$.  The average magnitude difference between the two catalogs for each galaxy and filter combination is less than 0.05. In Fig. \ref{fig:all_gal_colmag}, we plot the $M_{[4.5]}$ magnitudes vs $[3.6] - [4.5]$ colors for all the sources in our catalogs for each galaxy.  We also highlight the positions in the color-magnitude diagrams of the variables we present in \S\ref{sec:results}. Most of these variables have $[3.6] - [4.5] > 0$.

\begin{figure}
	\includegraphics[width = 0.5\textwidth]{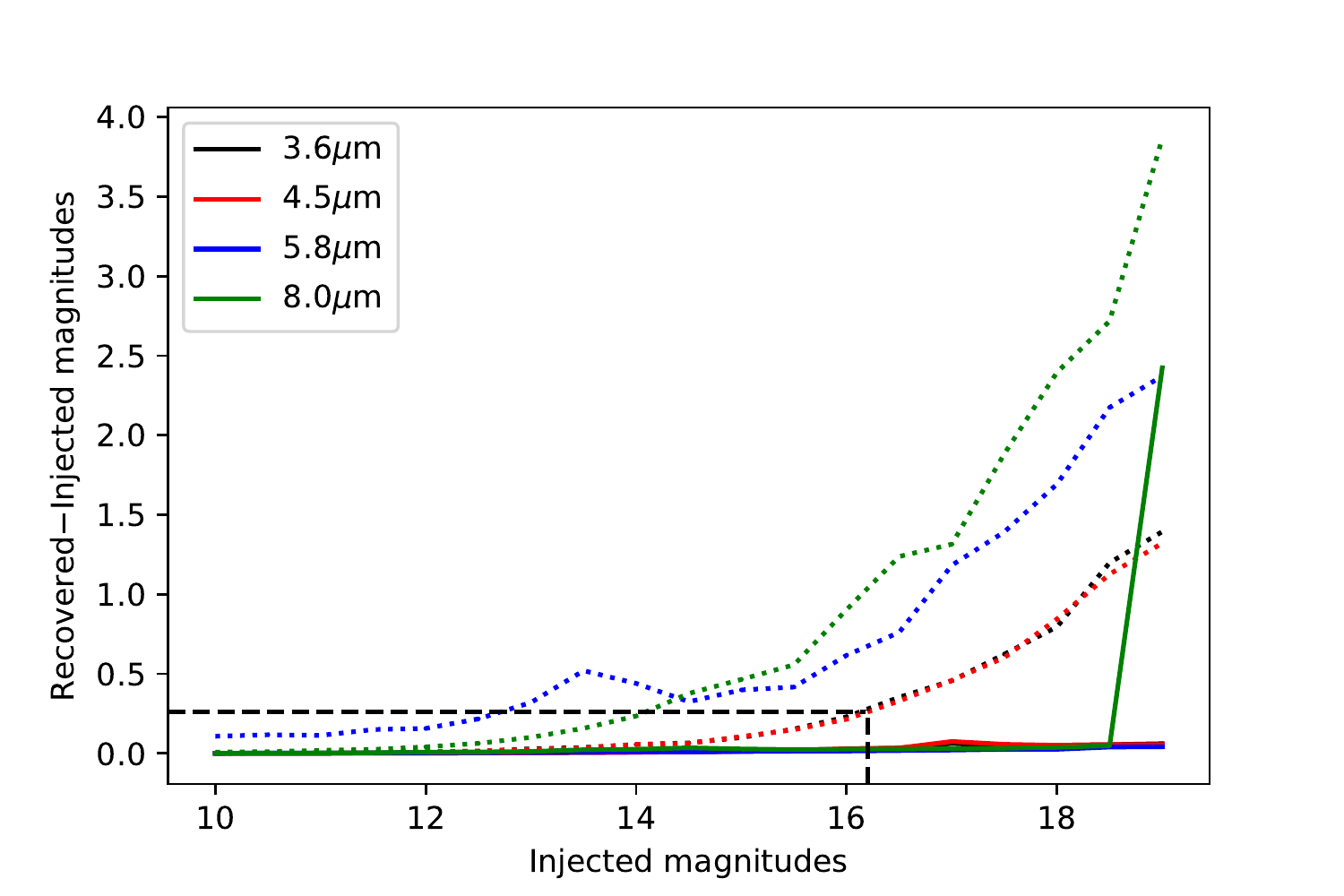}
    \caption{Artificial star test results for M101. The dotted lines indicate the behaviour of the 90$^{th}$ percentile of recovered sources and the solid lines indicate behaviour of 50$^{th}$ percentile of recovered sources. At $M_{[4.5]}=-13$ (marked in black dashed lines) for this galaxy, 90$\%$ of recovered sources have a bias less than 0.25. For the closer galaxies the bias should be even smaller, implying that our brightest variables do not suffer from large photometric biases. }
    \label{fig:art_test}
\end{figure}

It is possible that the magnitudes of the brightest sources are inflated due to biases in the photometric procedure. To check for this, we performed an artificial star test on M101 : the most distant and luminous galaxy in the sample that contains many of these bright variables. We used the derived PSFs to inject 1000 artificial stars with magnitudes ranging from 10$-$20 uniformly distributed into the 3.6, 4.5, 5.8 and 8.0 $\mu$m images of this galaxy. We perform the procedure described above to extract the magnitudes of these stars. We repeat this procedure 10 times and plot the behaviour of injected$-$recovered magnitudes for the 50$^{\mathrm{th}}$ and 90$^{\mathrm{th}}$ percentile of recovered sources in Fig. \ref{fig:art_test}. At $M_{[4.5]}=-13$ for this galaxy, 90$\%$ of recovered sources have a bias less than 0.15 mag. For the closer galaxies the bias should be even smaller, implying that our brightest variables do not suffer from large photometric biases. 

\subsection{Variable Source Identification and Photometry}
We utilize the list of variable sources identified from the image subtraction pipeline and visually vetted as described in \citet{Kasliwal17}.  In an effort to minimize contamination of the photometry from image subtraction artifacts and nearby variable sources, we use the smaller aperture sizes described in \S\ref{sec:refphot} to generate difference photometry (rather than the larger aperture sizes given in \citealt{Kasliwal17}).

The statistical uncertainties in the difference imaging photometry are much smaller than the true uncertainties arising from artifacts and other systematics. To more accurately reflect these uncertainties, for each epoch we add (in quadrature) the RMS of the reference-subtracted photometry of a grid of points within 7.2$\arcsec$ of the target to the statistical uncertainty.

We cross match the sources flagged by the SPIRITS pipeline to 3.6 and 4.5 $\mu$m reference catalogs using a matching radius of 1$\arcsec$ and retain the closest matching source.  The number of sources for each galaxy is given in Table \ref{tab:catalog_properties}.

\subsection{Period Fitting}
\label{sec:period_finding}
We simultaneously fit the 3.6 and 4.5$\mu\mathrm{m}$ light curves of all cross-matched sources with the {\sc gatspy} \citep{VanderPlas15,gatspycode} implementation of the Lomb-Scargle method \citep{Lomb1976,Scargle1982}.  Given the relatively sparse light curve sampling we restrict our periodicity search to sinusoidal signals.  We create an initial list of likely variables from the light curves that meet all of the following selection criteria: 
\begin{itemize}
\item null $\chi^{2}-$ reduced $\chi^{2}>3$, where null $\chi^{2}$ is the reduced $\chi^{2}$ for constant flux
\begin{itemize}
    \item ensures the periodic model is a significant improvement over a model with constant flux
\end{itemize}

\item best-fit period $<$ 0.8$\times$ light curve duration and best-fit period $<$ light curve duration $-$ 50 d
\begin{itemize}
    \item with poorly sampled data periods $\gtrsim$ the light curve duration are unreliable
\end{itemize}

\item best-fit period $>$ 250 d
\begin{itemize}
    \item periods must be significantly longer than typical cadences in order to avoid aliasing
\end{itemize}
\item best Lomb-Scargle score $>$0.8

\item (best Lomb-Scargle score) / (Second-best Lomb-Scargle score) $>$ 1.06
\begin{itemize}
    \item requiring the highest peak in the periodogram to be significantly higher than the 2nd highest peak eliminates many sources plagued by aliasing or poor fits
\end{itemize}
\item number of epochs of observation greater than $10$

\end{itemize}
We calculate phase-weighted mean magnitudes and peak to peak amplitudes for all the sources that meet these selection criteria. 

The number of observations and cadence varies between galaxies and between individual variables. Thus our confidence in the periods and amplitudes listed depends strongly on the details of the observations of the individual sources. We therefore aimed to assign a (subjective) quality to each variable identified, that is based on visual inspection of the light curves and periodograms $-$ gold:  very likely to be a periodic variable; silver: a possible periodic or quasi-periodic variable; bronze: not, or unlikely to be, periodic. However, given the large sample size of our sources, it is not feasible to do this for all variables. Hence, we limit this exercise only to the sources with $M_{[4.5]}$ brighter than $-12$, as these are brighter than the most luminous LMC AGB variables examined by \citet{Riebel10} and constitute the most interesting variables in our sample. Some of the interesting variables in the bronze classification are briefly described in Appendix A. 
At this point, we would like to draw attention to the fact that the process of identifying transients and variables involved members of the SPIRITS collaboration examining each new set of observations. These were presented to them as image subtractions, using earlier 3.6 and 4.5 $\mu$m images as references.  Individuals would identify ``sources'' as those of interest or junk. Depending on various factors including the density of earlier observations, the cadence, and the timing of the transient, some confusion between categories was inevitable. Some of the most interesting variables are in very crowded star forming regions, so isolating them is a challenge. Added to this is the fact that the cadence and the total number of observations are often not ideal for characterizing long-period variables. These facts combine to mean that this survey is incomplete and that in many cases we cannot definitively distinguish between periodicity and quasi-periodicity. In addition, crowding and limited sensitivity of IRAC to the distant galaxies in our sample are likely to inflate the magnitudes reported here by a small amount. Nevertheless, the results indicate the existence of a tantalizing population of large-amplitude infrared luminous sources with a range of properties, as discussed below.

\section{Results}
\label{sec:results}
We identify 417 candidate luminous LPVs in the galaxies in our sample and present their periods and phase-weighted mean absolute magnitudes,  $M_{[3.6]}$ and $M_{[4.5]}$, in Table \ref{tab:variable_catalog}.  We cross match these variables to the reference 5.8 and 8.0 $\mu$m images using a matching radius of 1$\arcsec$ and retain the closest matching source and find matches in both the 5.8 and 8.0 $\mu$m reference images for 160 of these variables.  

In the following we discuss possible origins of the variability of these SPIRITS sources in terms of pulsation (AGB variables), via interacting dusty winds in binary systems (WRC stars) or via orbital modulation, also in binaries. Other causes may contribute, e.g. eruptive variability, particularly if very long term monitoring suggests the variations are only quasi-periodic, e.g. LBVs.

\begin{figure*}
	\includegraphics[width = \textwidth]{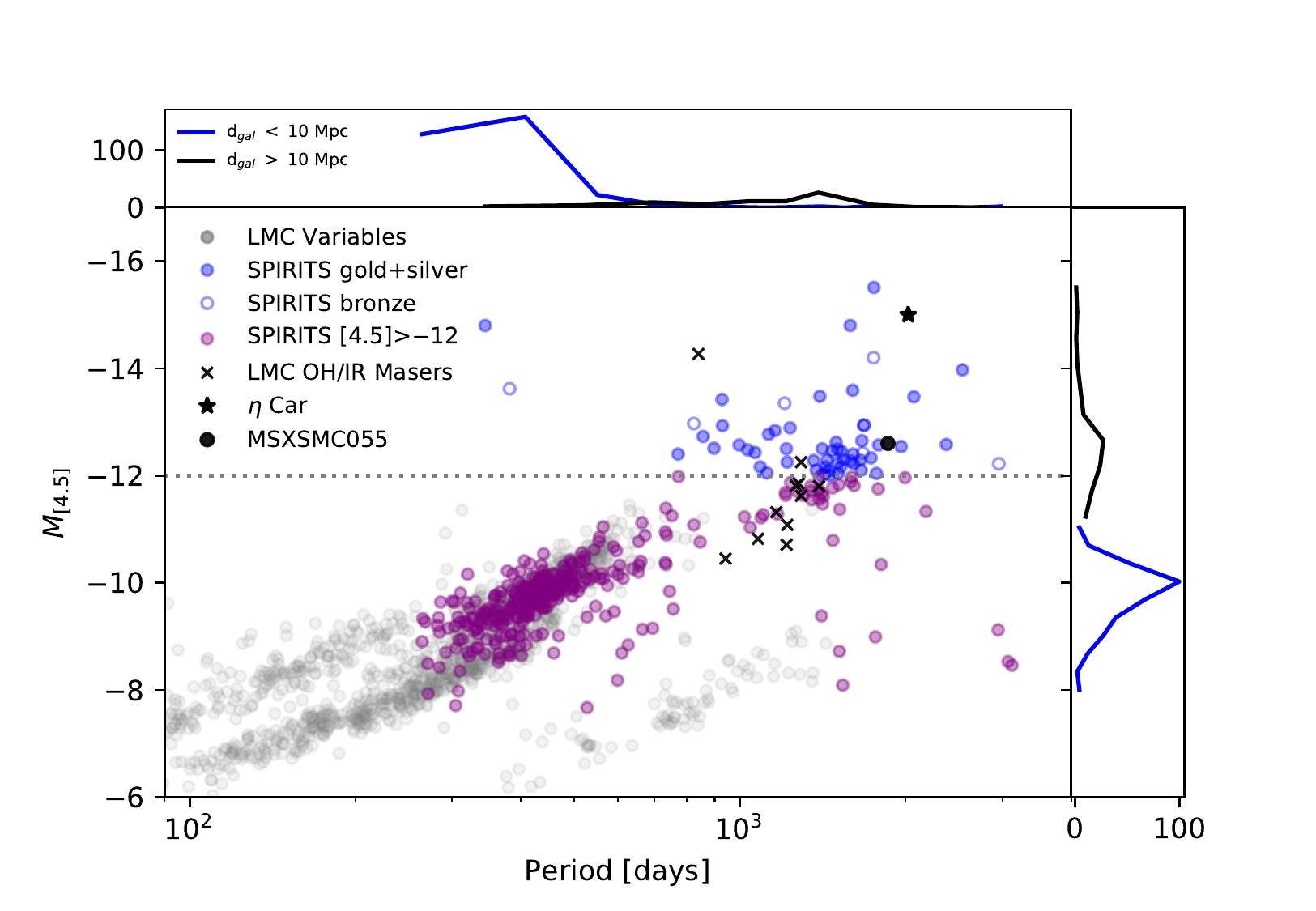}
    \caption{The period-luminosity diagram for the SPIRITS variables. The dotted line distinguishes the brightest variables ($M_{[4.5]}<-12$) which we discuss in \S\ref{sec:agb_candidates}. Many of the fainter variables ($M_{[4.5]}>-12$) are probably AGB stars pulsating in the fundamental or first overtone modes. Shown for comparison are the LMC AGB variables \citep{Riebel15} and LMC OH/IR sources \citep{Goldman17} (the most luminous of these, WOH G64, is a RSG), plus two individual objects discussed in \S\ref{sec:binaries}. The value of $M_{[4.5]}=-15.0$ used for $\eta$~Car is the mean of the 4.2 and 4.9 $\mu$m measurements from \citet{Price1983} and \citet{Ney1980}, respectively, and assumes a distance of 2300pc (although possibly it is more distant \citep{Davidson2018}). MSX-SMC-055 is both a candidate super-AGB star and a candidate TZO; its magnitude $M_{[4.5]}=-12.6$ is from \citet{Boyer+2011}. In the top and right panels, we plot the histograms of the periods and luminosities of SPIRITS variables respectively. The blue histograms represent the variables in nearby ($d<10$Mpc) galaxies, and the black histograms represent variables in distant ($d>10$ Mpc) galaxies. As is evident, we have a bimodal distribution of variables, where the low period $-$ lower luminosity variables preferentially belong to the nearby galaxies, and the longer period, more luminous variables belong to the distant, more massive galaxies. This apparent bimodal distribution is a consequence of our limited sensitivity, in that we are not sensitive to the low luminosity variables from the distant massive galaxies, and the nearby galaxies have much lower stellar masses, and are thus unlikely to host the long period luminous variables. }
    \label{fig:pldiagram}
\end{figure*}

We plot the period luminosity relation for the sources in our sample, using the derived periods and their phase weighted mean magnitudes in Fig. \ref{fig:pldiagram}. 
The sequences corresponding to the fundamental and first overtone modes of evolved variables in the LMC \citep{Riebel10} are populated  in the diagram. The variables in the nearby galaxies in our sample closely match luminosities and periods of the LMC variables (for periods longer than our 250 d threshold and absolute magnitudes brighter than -8).  Along with these, we also find variables that are more luminous and have longer periods than the LMC variables and lie in a previously largely empty region of the period-luminosity diagram. We note that there is a scarcity of variables with periods between 650 $-$ 1200 days, and $M_{[4.5]}$ between $-11$ and $-12$. This gap is most likely a product of our limited sensitivity, and we discuss it in \S\ref{sec:agb_candidates}.

\begin{figure*}%
	\includegraphics[width = \textwidth]{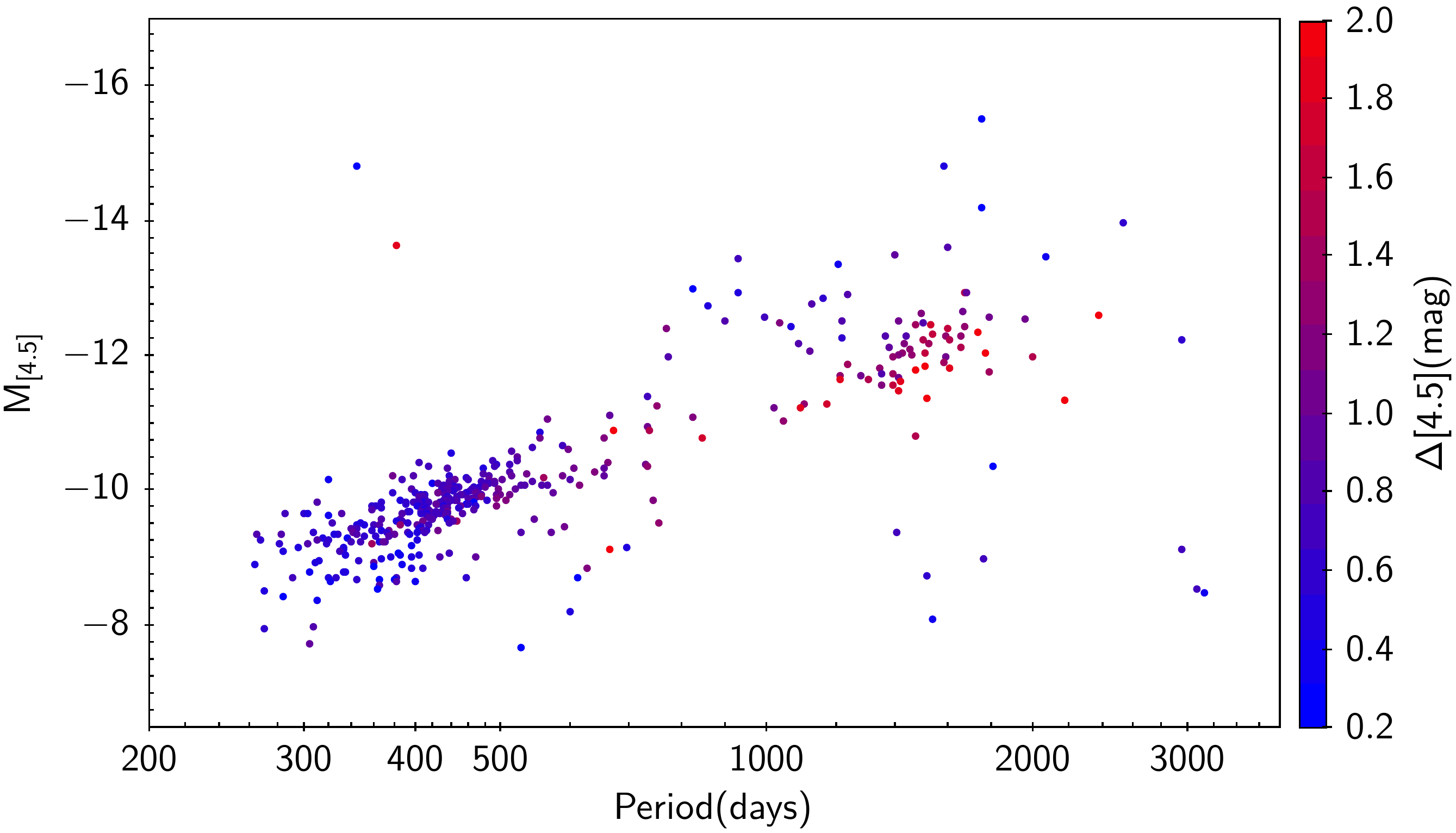}
    \caption{Period-luminosity diagram of SPIRITS sources, color coded by their amplitudes. A majority of our low luminosity AGB candidates have relatively small amplitudes, from 0.5 to 1 mag. The higher amplitude sources are found predominantly at higher luminosities and longer periods. The group around $\rm P \sim 1500$ days and $M_{[4.5]}\sim -12$, which we think are relatively massive AGB stars, have particularly large amplitudes.}
    \label{fig:pldiagramamp}
\end{figure*}

In Fig. \ref{fig:pldiagramamp}, we plot the SPIRITS variables color coded by their peak-to-peak amplitudes of variability. We note that a majority of our lower luminosity sources have relatively low amplitudes, from 0.5 to 1. The higher amplitude sources are predominantly found at higher luminosities and longer periods and in particular within the group that are probably high mass AGB stars (see \S \ref{sec:brightest_sources}).

\begin{figure}
	\includegraphics[width = 0.5\textwidth]{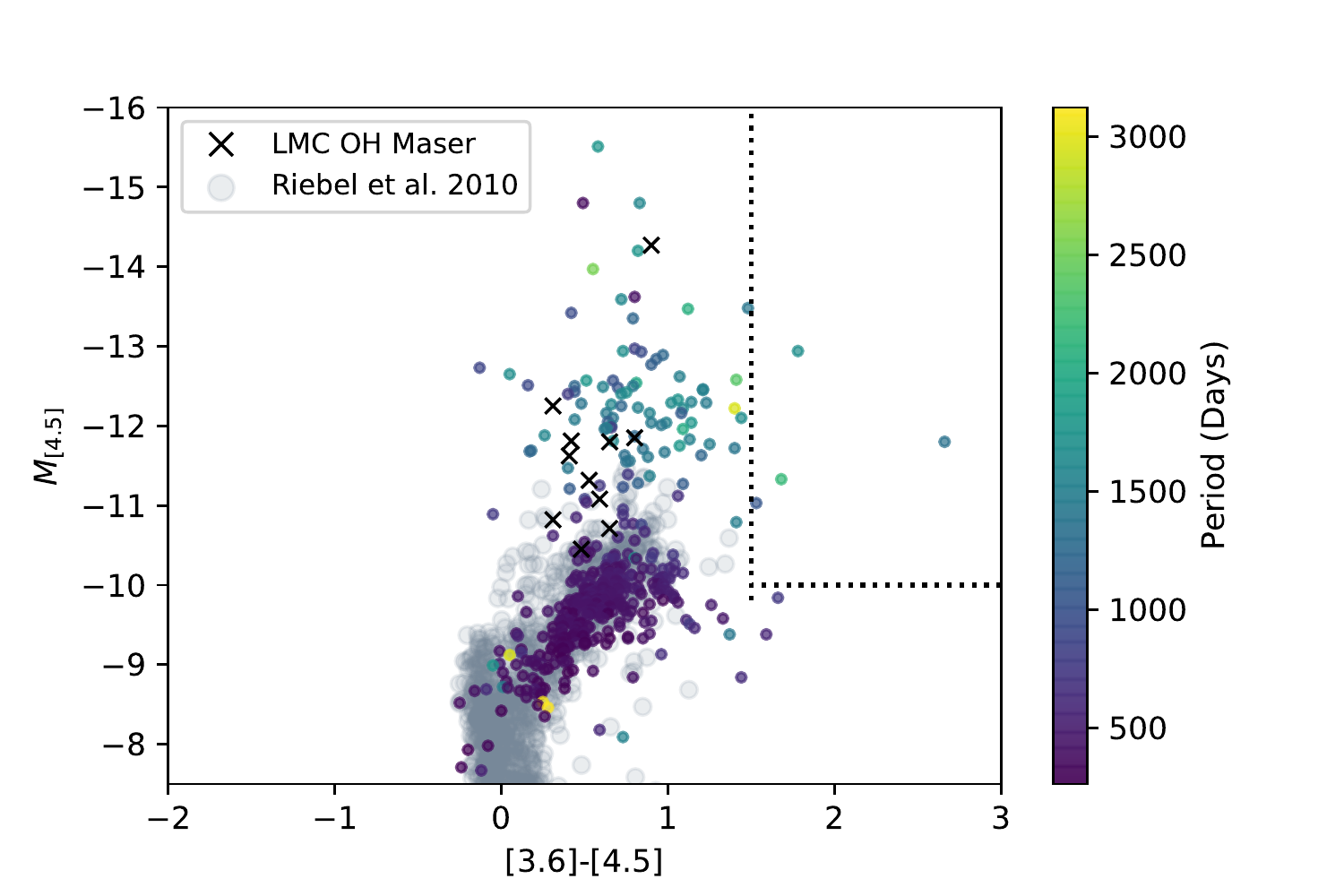}
    \caption{Mean $M_{[4.5]}$ vs $[3.6]-[4.5]$ color magnitude diagram for SPIRITS variables, along with LMC AGB variables (smaller red points). The shorter period and lower luminosity variables in our sample coincide with the LMC variables, and hence are most likely AGB stars. The more luminous and longer period variables are clearly separated from the AGB region in the diagram. These sources have colors ranging from $-0.5$ to 1.5. Some of them occupy the same part of the colour-magnitude diagram as LMC OH Masers (black crosses).  We also indicate the SN2008s-like region of \citet{Thompson09} ($M{[4.5]}<-10$  and $[3.6] - [4.5] > 1.5$ mag). There are three variable sources in this extremely red region.}
    \label{fig:cmd36-45}
\end{figure}

\begin{figure}
	\includegraphics[width = 0.5\textwidth]{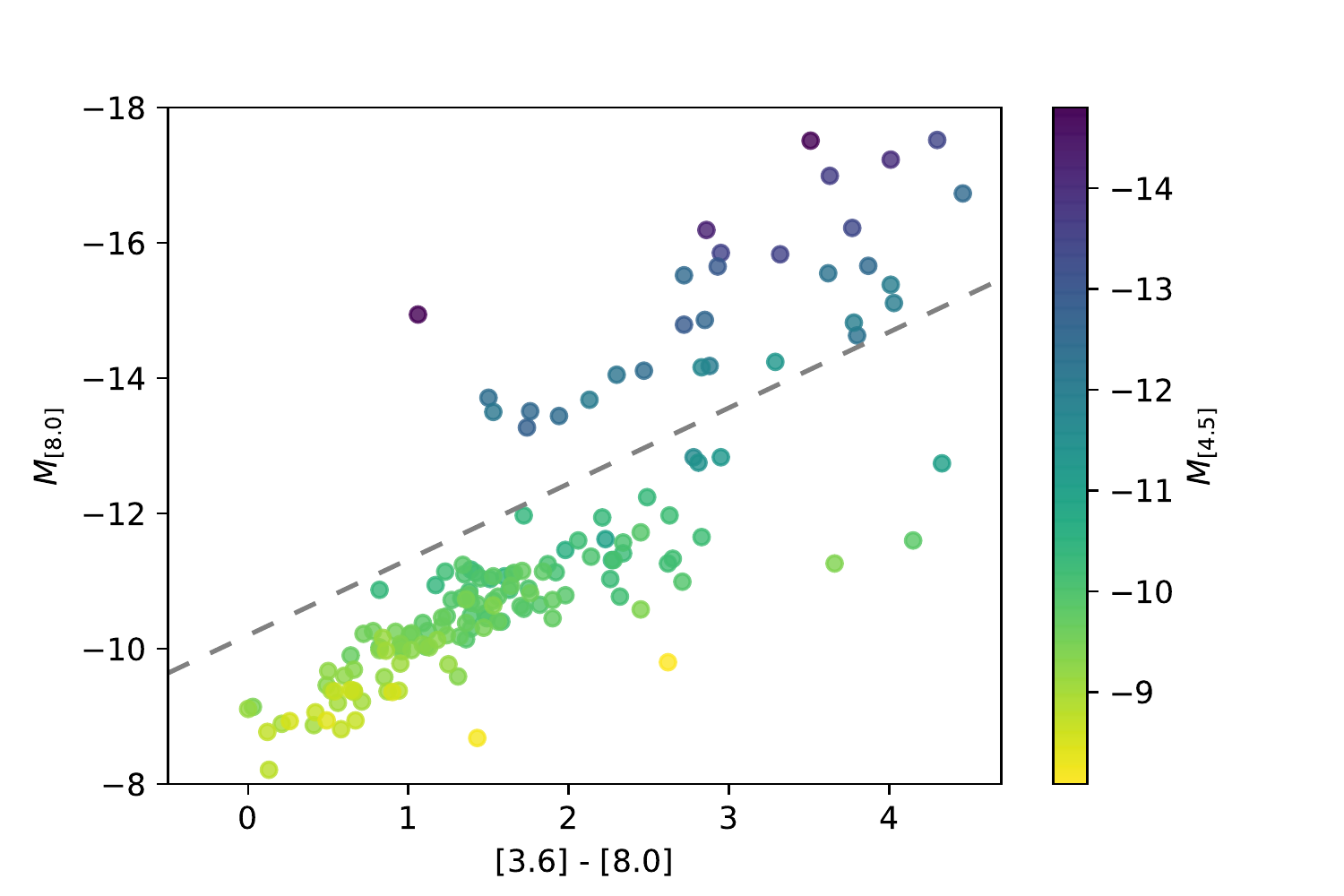}
    \caption{Reference [8.0] vs $[3.6] - [8.0]$ CMD for the SPIRITS variables, color coded by their $M_{[4.5]}$. We note that our AGB candidates exhibit a sequence, while the luminous ([4.5] brighter than $-12$) candidates lie on a different, more luminous sequence. The gray line roughly indicates the separation of these two sequences. However, most of the luminous sources belong to clusters or are near HII regions. This coupled with the fact that the PSF is large at 8.0$\mu$m could inflate the brightness and colors of these sources. }
    \label{fig:cmd36-80}
\end{figure}

\begin{figure}
	\includegraphics[width = 0.5\textwidth]{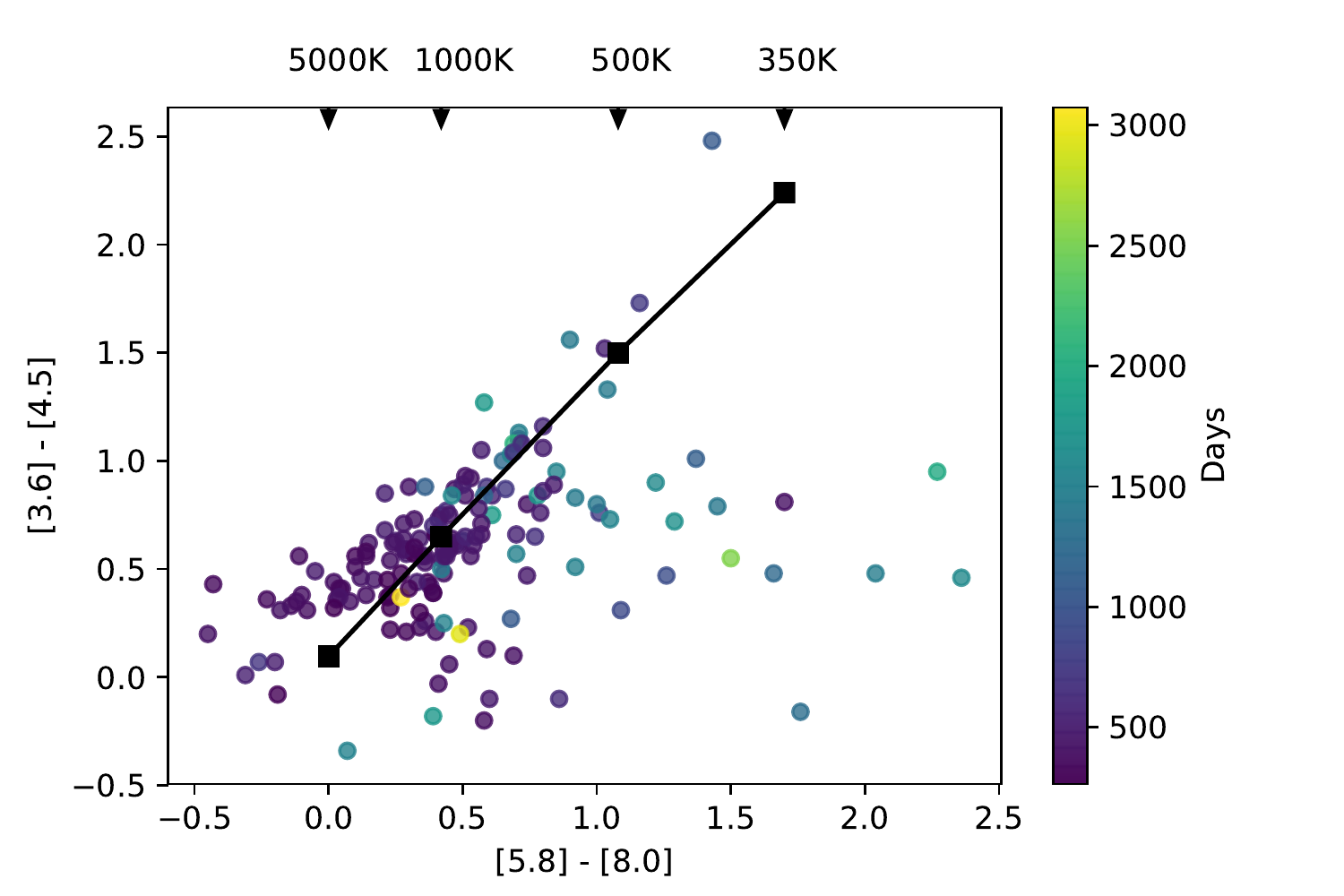}
    \caption{Color color diagram for SPIRITS sources. The black line and squares denote the expectation for a blackbody of temperature 5000, 1000, 500 and 350K. The sources that lie below the blackbody line have an IR excess due to the dust enshrouding them. A colder dust shell pushes a star to the bottom right of this diagram, with larger $[5.8]-[8.0]$ colors }
    \label{fig:colcol}
\end{figure}

We also plot the $M_{[4.5]}$ vs $[3.6]-[4.5]$ color magnitude diagram of the SPIRITS variables in Fig. \ref{fig:cmd36-45}, color-coded by their period. Overplotted are the variables from the LMC reported by \citet{Riebel15}. We find a large group of variables with periods in the range of 250 - 750 days, which coincides with the AGB region of the LMC. The longer period variables are more luminous than these and have colors between $-0.5$ and 1.5 . Also marked in the color-magnitude diagram is the region identified by \citet{Thompson09} ($M_{[4.5]} <-10$  and $[3.6] - [4.5] > 1.5$ mag) containing progenitors of SN2008S-like events. We also plot the $M_{[8.0]}$ vs $[3.6]-[8.0]$ color magnitude (Fig. \ref{fig:cmd36-80}) and $[3.6] - [4.5]$ vs $[5.8] - [8.0]$ color color (Fig. \ref{fig:colcol}) diagrams for the sources that find matches in the 5.8 and 8.0 $\mu$m reference images. We note that in Fig. \ref{fig:cmd36-80}, our AGB candidates exhibit a sequence, while the luminous ($M_{[4.5]}$ brighter than $-12$) candidates lie on a different, more luminous sequence. In the color$-$color diagram (Fig. \ref{fig:colcol}), we also plot the expectation for a blackbody at temperatures of 5000, 1000, 500 and 350K. The diagram suggests that the long period sources in our sample are extremely cold and have an IR excess due to large amounts of dust enshrouding them. However, most of the luminous sources belong to clusters or are near HII regions. This coupled with the fact that the PSF is large at 8.0 $\mu$m could inflate the brightness and colors of these sources. 

\subsection{Comparison of SPIRITS and DUSTiNGS}
\label{sec:dustings}
\begin{figure}
	\includegraphics[width = 0.5\textwidth]{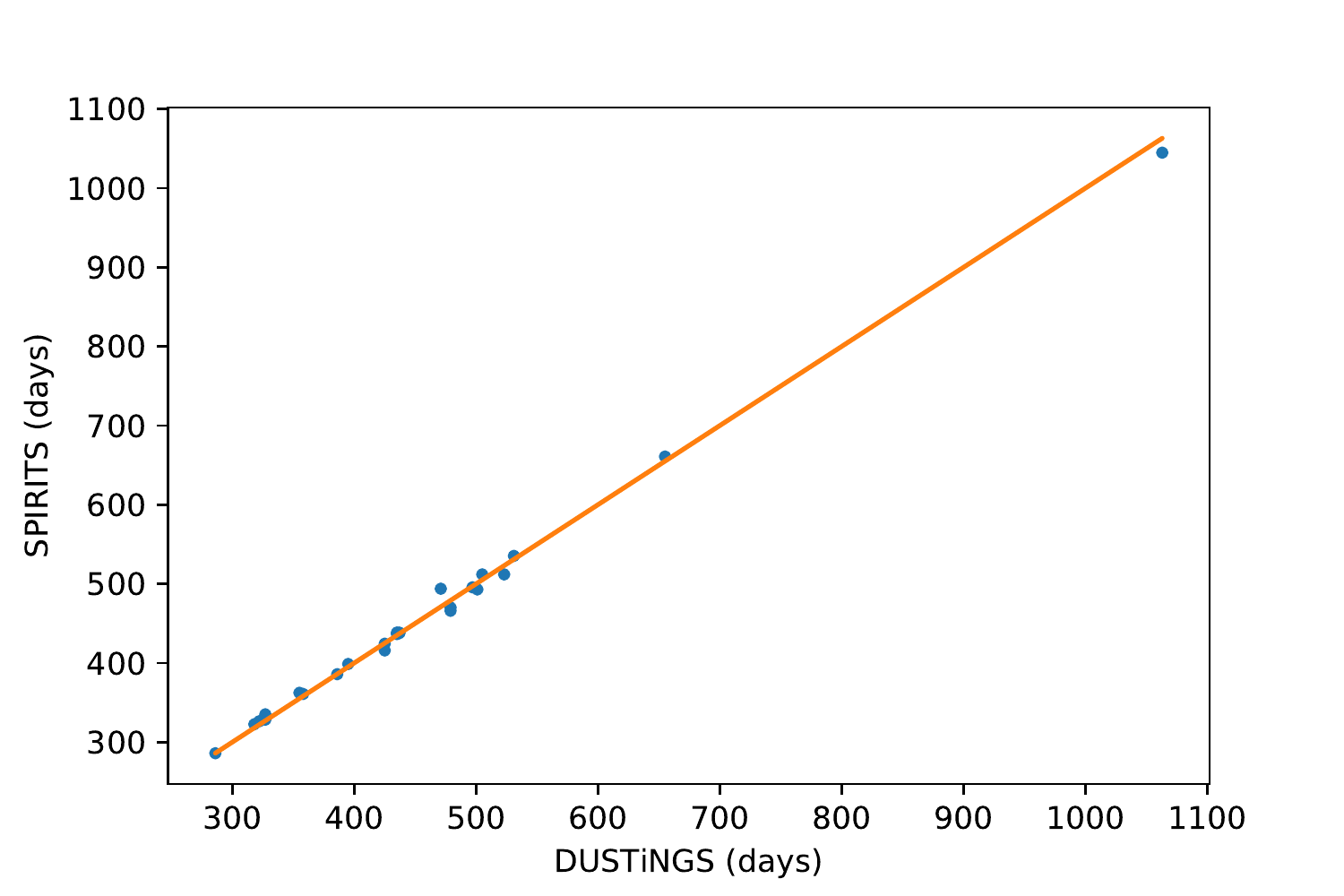}
    \caption{Comparison of SPIRITS and DUSTiNGS periods.}
    \label{fig:dustings_periods}
\end{figure}

\begin{figure}
	\includegraphics[width = 0.5\textwidth]{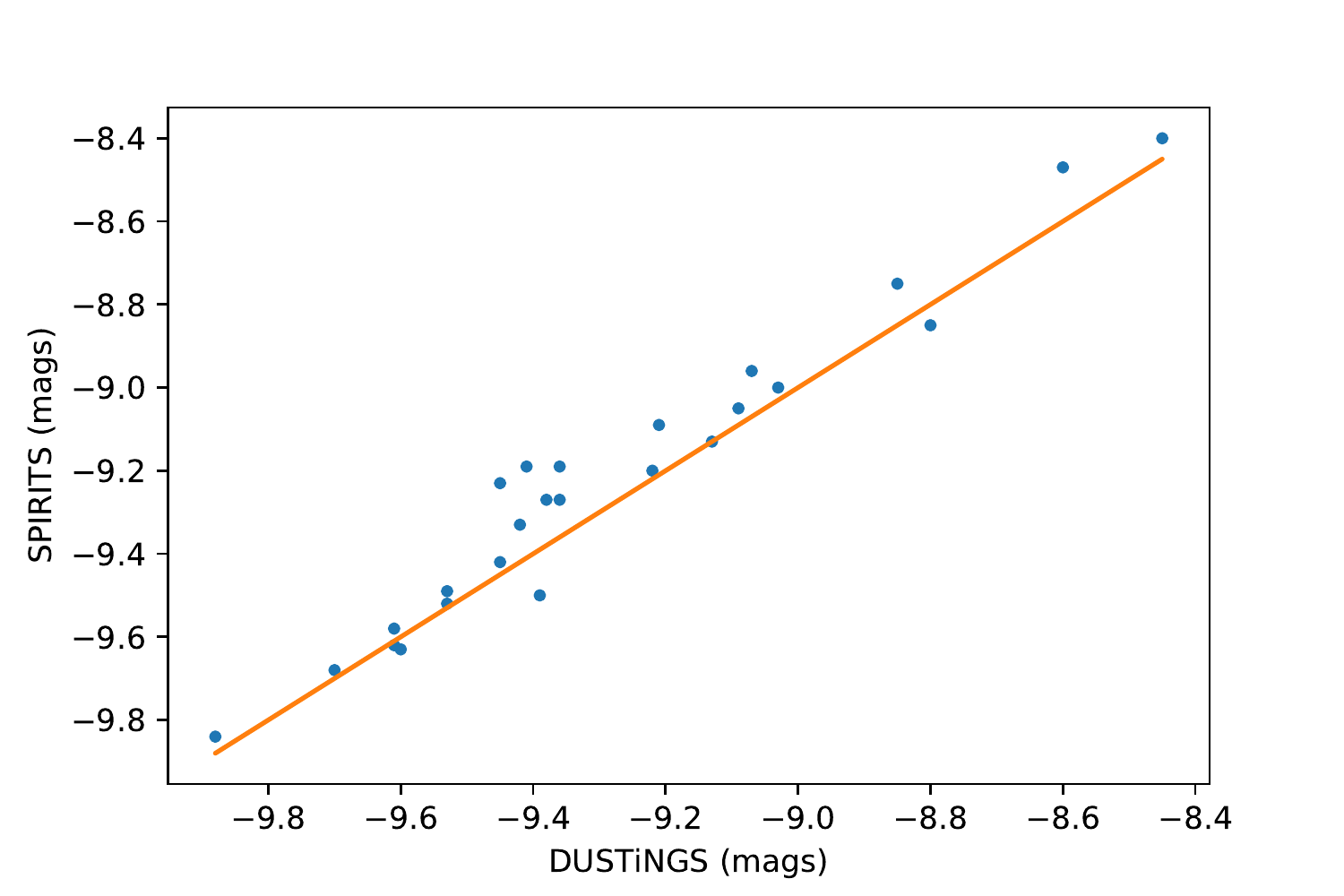}
    \caption{Comparison of SPIRITS and DUSTiNGS 3.6 $\mu$m absolute magnitudes.}
    \label{fig:dustings_mags}
\end{figure}

\begin{figure}
	\includegraphics[width = 0.5\textwidth]{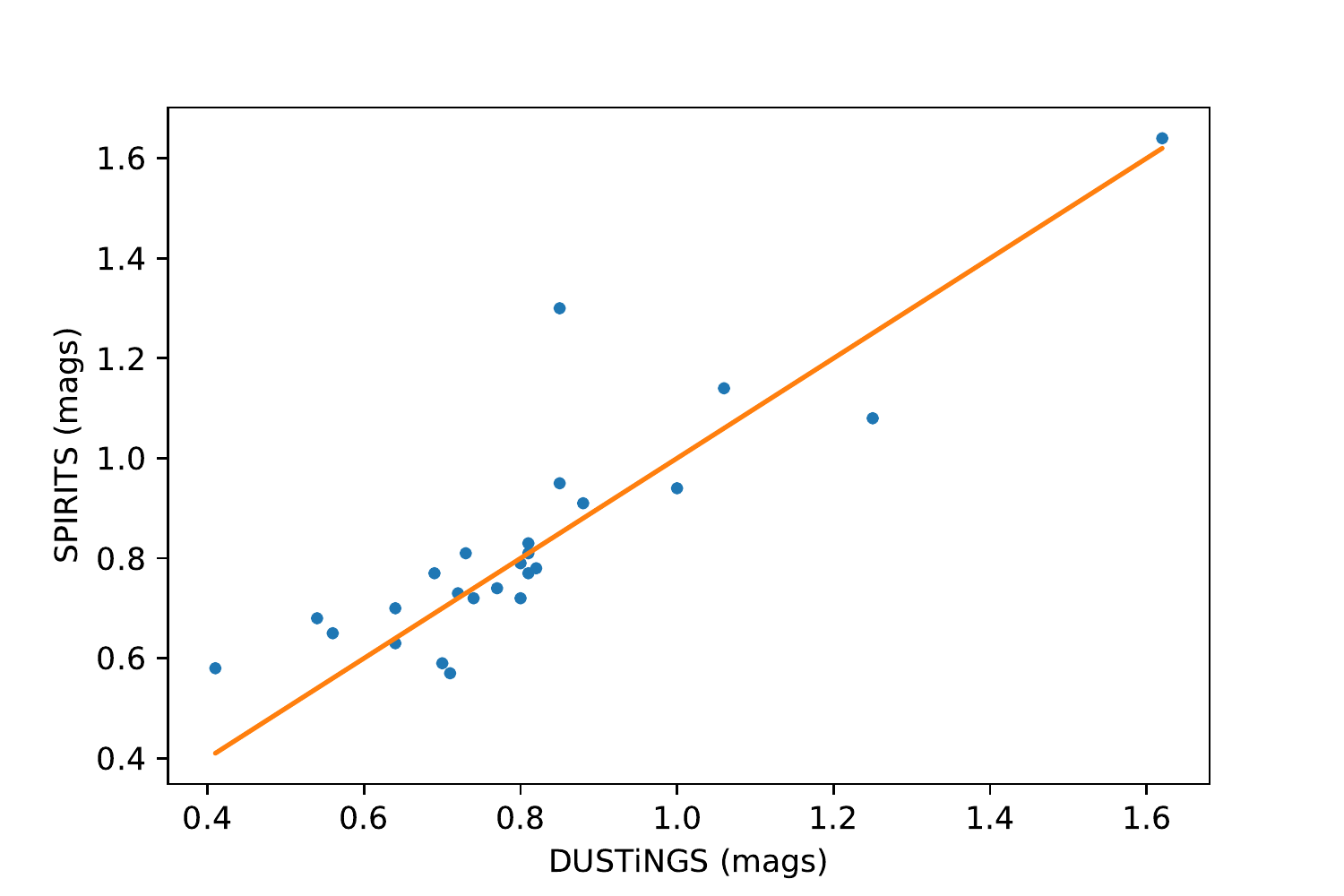}
    \caption{Comparison of SPIRITS and DUSTiNGS 3.6 $\mu$m amplitudes.}
    \label{fig:dustings_amps}
\end{figure}

In this section we compare our variable catalog with the four galaxies in the DUSTiNGS survey recently analyzed by Goldman et al, (in prep). The DUSTiNGS survey identifies 62 AGB candidates in the four galaxies we have in common (NGC147, NGC185, IC1613 and WLM), while we identify 95 variables in these galaxies. 25 sources are common between the SPIRITS and DUSTiNGS lists. It is possible that some of the DUSTiNGS variables were missed by SPIRITS during the process of human scanning (see \S\ref{sec:period_finding}). Figures \ref{fig:dustings_periods}, \ref{fig:dustings_mags} and \ref{fig:dustings_amps} compare our periods, [3.6] absolute mean magnitudes and [3.6] amplitudes with corresponding DUSTiNGS values for these 25 variables. The SPIRITS periods and magnitudes agree well with the DUSTiNGS values, with an average scatter of 9 days and 0.1 mag respectively, while the amplitudes have a slightly larger scatter of 0.13.

\subsection{AGB Candidates}
\label{sec:agb_candidates}
We identify 359 variables in our sample with 4.5 $\mu$m mean absolute magnitudes fainter than $-12$ (which is close to the maximum luminosity of the LMC AGB sample examined by \citet{Riebel10}). Nearly all of these lie along the known period luminosity relation for AGBs. These sources predominantly belong to the nearby dwarf galaxies and are most likely pulsating AGB stars from these galaxies. Like the LMC variables, most of the sources in our sample also fall where we would expect for the fundamental or first overtone pulsators in the period-luminosity diagram. However, as mentioned in \S\ref{sec:period_finding}, we do not visually inspect the light curves of these candidates. Thus it is possible that some of these sources do not have well defined periods.

Most of these variables have periods around $250 - 750$ days, while a handful of them are oscillating with longer periods around and larger than 1000 days.
The fundamental pulsators have absolute magnitudes between $-8$ and $-11$. As a representative of this group, we present the phase-folded light curve of SPIRITS~14va (Fig. \ref{fig:14va}) in the galaxy WLM, with a period of 511 days and a mean 4.5 $\mu$m magnitude of $-10.26$.


The longest period variables with $M_{[4.5]}>-12$ have periods ranging from 1400 up to 3200 days. SPIRITS~15acg (Fig. \ref{fig:15acg}) in NGC~185 belongs to this category, with a period of 1407 days, as do SPIRITS~14ct (Fig. \ref{fig:14ct}) in NGC~6822, with a period of 2947 days, and SPIRITS~15aci (Fig. \ref{fig:15aci}) in NGC~185. Some of these are known C-stars, which possibly have long secondary periods or erratic variations; SPIRITS~14bjc \citep{Letarte2002} in NGC~6822 and SPIRITS~15aci \citep{Nowotny2003} are examples of this type. Other variables are almost certainly Quasars (or other Active Galactic Nuclei), SPIRITS~14ary ($M_{[4.5]} = -10.34$  period = 1806 days), nominally in IC~1613, would be one example of these \citep{Flesch2017}. 

In the $M_{[4.5]}$ vs $[3.6] - [4.5]$ color magnitude diagram (Fig. \ref{fig:cmd36-45}), the AGB candidates having $M_{[4.5]}$ fainter than $-11$  form a group and trace the AGB of the LMC variables color magnitude diagram. The brighter AGB candidates in this sample show more scatter, but there are no significant deviations from the trends exhibited by the LMC AGB variables. The three long period candidates ($>2000$ days) have $[3.6]-[4.5] < 0.5$  and occupy the same region as shorter period AGB candidates, supporting the idea that the long periods found from the Spitzer data may be secondary. 

The reddest and brightest sources among these are referred to as extreme AGB (eAGB) variables. \citet{Thompson09} identify eAGB stars in M33, and look for progenitors of SN2008S and NGC300-like transients in these galaxies. They note that  reasonable criteria for eAGB stars to be identified as analogs of progenitors of these transients are $M_{[4.5]} < -10$, $[3.6]-[4.5] > 1.5$, and small amplitude  variability (amplitude $<0.3$ mag). There are four periodic variables in this extremely red region of the CMD, but all have amplitudes larger than 0.9 mag. 

There is a tantalizing paucity of stars on the extrapolated PL relation with periods between 650 and 1200 days.  However, this gap may be an artifact of the differing sensitivities of our two-tiered galaxy sample.  While our cadence and light curve durations maintain sensitivity to periods between 650 and 1200 days, the survey is not sensitive to variables fainter than $M_{[4.5]} \sim -11$ (see Fig. \ref{fig:all_gal_colmag}) in the massive, more distant galaxy sample.  Since the stellar mass in the massive galaxy sample is $\sim$1000$\times$ that of the nearby dwarf sample, the luminous ($M_{[4.5]}<-12$) variables are clearly much rarer than the fainter variables identified in the nearby sample.  
If variables in the apparent period--luminosity gap are similarly rare as the luminous variables the observed gap would arise because the nearby dwarfs may only harbor a few of them and they are too faint to be detected in the massive, more distant galaxy sample.

\subsection{Brightest sources}
\label{sec:brightest_sources}
In this section we discuss the 58 sources in our sample with mean absolute 4.5~$\mu$m magnitudes brighter than $-12$. We classify 24 of these sources as ``gold'', 26 as ``silver'' and 8 as ``bronze'' variables (Table \ref{tab:variable_catalog}).

Eleven of these sources have mean $M_{[4.5]}$ luminosities brighter than $-13$. Of these, 9 have long periods ($> 900$ days):  SPIRITS~15js, SPIRITS~15jt, SPIRITS~16do and SPIRITS~16po in M83, SPIRITS~14apu, SPIRITS~14apz and SPIRITS~15pk in M101, SPIRITS~15ty in NGC~6946 and SPIRITS~18ec in M51. Short periods were found for two sources: 344 days for SPIRITS~14aue (Fig. \ref{fig:14aue}) in IC~342 and 381 days for SPIRITS~14axv (SN2005at) in NGC~6744. SPIRITS~14apu (Fig. \ref{fig:14apu}), the most luminous source in the sample, has mean absolute $M_{[4.5]}=-15.51$  a period of 1752 days, a [4.5] amplitude of 0.25 mag and a $[3.6]-[4.5]$ color of 0.58. It is one of the M101 cluster sources discussed below. Two of these extra-luminous sources have periods longer than 2000 days : SPIRITS~16do (2071 days, Fig. \ref{fig:16do}) and SPIRITS~18ec (2538 days, Fig. \ref{fig:18ec}). Seven of these luminous long period ($>900$d) variables have matches in the 5.8 and 8.0 $\mu$m catalogs. These bright sources have reference 8.0 $\mu$m magnitudes brighter than -16. They are extremely red with $[3.6]-[4.5]$ colors between 0.42 and 1.48  and reference $[3.6] - [8.0]$ colors between 3.0 and 4.75.

We find 47 sources with mean absolute 4.5 $\mu$m magnitudes between $-12$ and $-13$. As a representative of this sample, we present the folded light curve of the source SPIRITS~14apq (Fig. \ref{fig:14apq}), which has a period of 1606 days and a mean 4.5 $\mu$m absolute magnitude of $-12.40$. The periods of these sources range from 770 days to 2956 days. Most of them are extremely red, with $[3.6] - [4.5]$ colors between 0.4 and 1.5. Eighteen of these sources have counterparts in 5.8 and 8.0 $\mu$m images. Their reference [8.0] magnitudes range from $-13.5$ to $-17$ and their reference $[3.6] - [8.0]$ colors lie between 1.5 to 4.5. 

There is a group of variables that overlap the
longest period LMC OH/IR stars in Fig.~\ref{fig:pldiagram}. These have periods between 1300 and 1800 days and absolute $4.5\mu$m magnitudes between $-11.5$ and $-12.5$. We represent them by the phase folded light curve of the source SPIRITS~15kb (Fig. \ref{fig:15kb}) in the galaxy M83. SPIRITS~15kb has a period of 1458 days and a mean absolute 4.5 $\mu$m magnitude of $-12$. This group of sources has large amplitudes, larger in general than than the even brighter ($M_{[4.5]}<-13$) sources. These are probably AGB stars with massive progenitors.

\begin{figure*}
	\includegraphics[width=\textwidth]{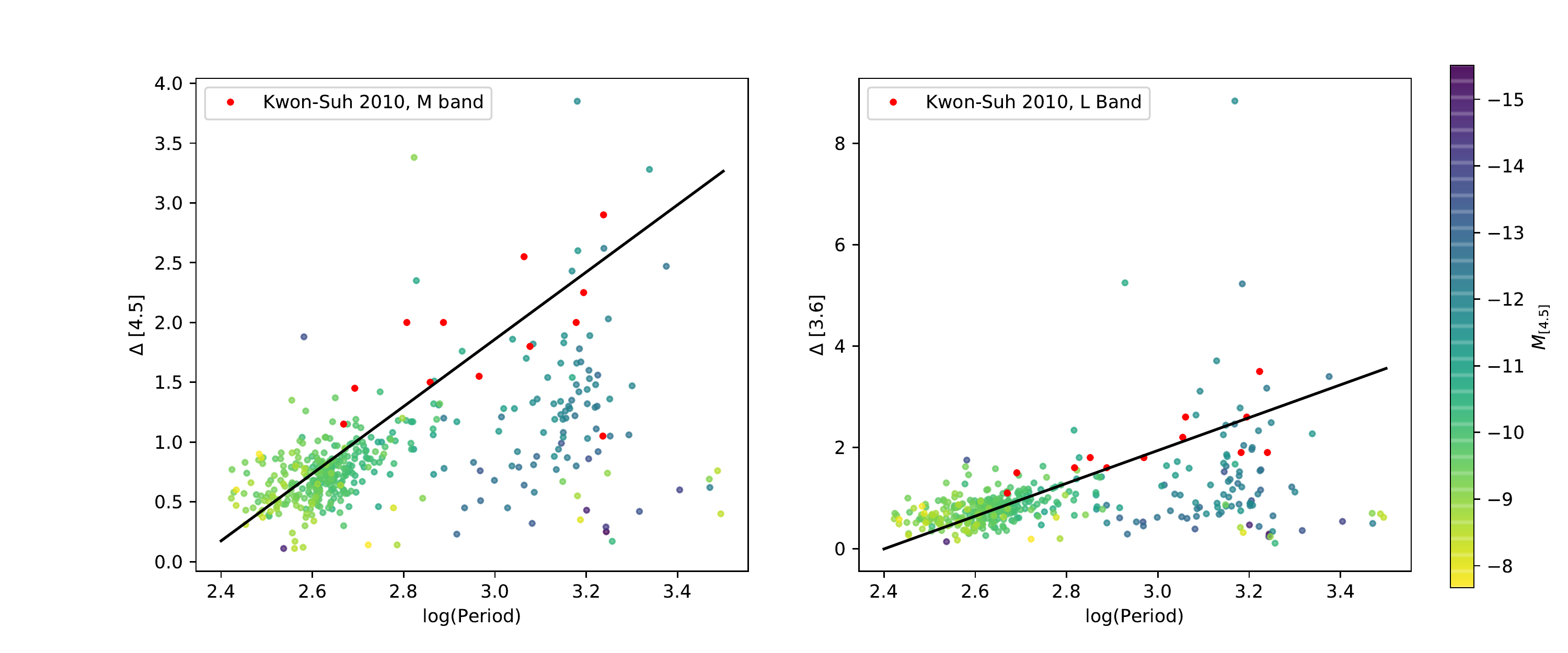}
    \caption{Here, we compare our variables with the $L$ and $M$ band amplitude-period relation (black solid line) derived by \citet{KwonSuh2010} for Galactic OH/IR Masers (red dots). While our lower luminosity variables are found in the low period- low amplitude region, the higher luminosity variables show a large spread. Some of these could have circumstellar maser emission.}
    \label{fig:kwon}
\end{figure*}

\citet{KwonSuh2010} present an empirical $M$- and $L$-band amplitude-period relation using 12 galactic OH/IR stars. We compare the $M_{[4.5]}$ and the $M_{[3.6]}$ amplitude-period relation for SPIRITS variables to the $M$- and $L$-band relations respectively in Fig.~\ref{fig:kwon}. While our lower luminosity variables are found in the low period/low amplitude region, the longer period variables show a huge spread in amplitude. We note that MSX SMC055, the super-AGB and TZO candidate discussed in \S 3.3.2, has a modest amplitude ($\Delta L \sim 0.7$ mag) compared to the Galactic OH/IR stars.  At least some of the high luminosity variables are candidates for OH/IR Masers. 

Some of the brightest SPIRITS sources have been identified in other catalogs and have Two Micron All Sky Survey \citep[2MASS;][]{Milligan96}, Wide-field Infrared Survey Explorer \citep[WISE;][]{Wright2010} and Panoramic Survey Telescope and Rapid Response System \citep[PanSTARRS;][]{Kaiser2002} photometry. 
The last column of Table \ref{tab:variable_catalog} contains references for specific sources, where they are associated with clusters, HII regions, SN remnants etc.
\subsubsection{Variables in clusters}
\label{sec:clusters}
A large fraction of the most luminous variables ($M_{[4.5]}$ brighter than $-12$) in M101 and M83 may be associated (some in, others nearby) with star clusters. Some of the luminous variables not obviously associated with clusters are very close to HII regions (e.g., 18ec in M51, \citet{Lee2011}), Giant Molecular Clouds (e.g., 15ty in NGC6946, \citet{Donovan_Meyer2013}) and/or SNe remnants (e.g., 15jt in M83, \citet{Winkler2017}), and it is not possible to determine if they are associated with clusters as well. Most of the other galaxies with luminous variables have not been surveyed with HST, so no comparison can be made. The brightest sources, those with $M_{[4.5]}$ brighter than $-13$, are probably RSGs. So we expect that most of them will be found in clusters, and that these clusters would be unresolved by \emph{Spitzer}.  Accordingly the magnitudes of these sources are likely somewhat inflated by the light of the rest of the cluster. However, four of these 11 brightest variables vary by more than 50\%, meaning their phase-weighted mean magnitudes are dominated by the variable and not the cluster light.

\citet{Grammer2013} discuss the massive star population in M101 from HST photometry.  Ten of our twelve very luminous variables in M101 are coincident with large numbers of their massive stars, with between 13 and 155 stars within a radius of three arcsec of the Spitzer position. There are only a few red supergiants ($V<24$, $V-I>1.2$) amongst these, but we would not expect very dusty sources to show up at HST wavelengths.  There is no optical counterpart in the HST images at the positions of the other two sources (SPIRITS~14bco and SPIRITS~18ae) in M101. \citet{Grammer2015} have examined some of the massive stars in more detail, obtaining spectra and looking for variability. They found various hot and warm supergiants, LBV and Wolf-Rayet (WR) candidates. None of their variables are close to the sources we discuss.

\citet{Silva-Villa2011} discuss clusters in M83 and three other galaxies using HST photometry. As indicated in Table \ref{tab:variable_catalog} five variables may be associated with one or more suspected or accepted cluster in M83, although not as closely associated as the M101 sources are with their clusters.  \citet{Silva-Villa2011} suggest these M83 clusters are of intermediate age,  $10^8 <\rm age<10^9 \mathrm{yr}$ and intermediate total mass, $10^{3.3} <M<10^{4.4}\ M_{\odot}$, noting that it is difficult to estimate age or mass for these clusters.  \citet{Ryon2015} also discuss clusters in M83, three of which are among our sample, two in common with \citet{Silva-Villa2011}. They find slightly younger ages, $<10^{8.5}$ yr, and comparable masses for the three of interest. \citet{Larsen2011} presents colour-magnitude diagrams of few individual clusters including NGC5236-F1-3 in M83 which coincides with SPIRITS~15zg. They derive an age of 10$^{7.5}$ yr for this cluster. We note that at the distance of M83 a star traveling at the modest velocity of $30\, km s^{-1}$ would take only about $2\times 10^6$ yr to travel three arcsec and could therefore have moved a long way from the birth site in $10^7$ to $10^8$ yr. A more detailed study is required to establish if these stars were flung out of clusters some distance from where they now are.

If the ages mentioned above for M83 are correct we would not expect to see red supergiants or very luminous AGB stars, which will not be older than about $10^8$ yr. The clusters in M101 could certainly include RSGs and/or super-AGB stars. Interacting binaries are of course possible, even likely, in both galaxies.

\subsubsection{Binaries}
\label{sec:binaries}
It seems almost certain that some of these luminous variables will be interacting binaries. A comprehensive discussion of the numerous alternatives is beyond the scope of this discovery paper, but we do briefly examine a few possibilities. We start with $\eta$~Car, which is a very massive star (around $100\ M_{\odot}$) in a highly eccentric binary system \citep{Damineli1996}, which varies on many timescales and emits at every wavelength \citep[see, e.g.,] [and references therein] {Davidson2012}. It is often classified as an LBV, although it differs significantly from other LBVs (e.g., most of the others are blue and very few are known binaries). Even if such massive stars are very rare, $\eta$ Car is extremely red (its mass loss rate exceeds $10^{-3}\ M_{\odot} yr^{-1}$ \citep[e.g.,][]{Hillier2001}) and extremely luminous and similar objects should be anticipated in a mid-infrared survey of galaxies with large populations of very young stars; although confusion within star-forming regions will probably present challenges. $\eta$ Car has an $L$ amplitude of about 0.4  \citep{Whitelock2004} and is shown in Fig. \ref{fig:pldiagram} at the orbital period of 2023 days. Although we note that the character of the variations is somewhat different from those of most of the SPIRITS variables, in that the $\eta$~Car variation is concentrated around the time of periastron. Other LBVs are much less luminous in the infrared, although some have exhibited quasi-periodicity on the timescale under discussion.


A majority of late-type Carbon-rich Wolf-Rayet (WCL) binaries are efficient dust producers ($\dot{M}\gtrsim 10^{-8}$ M$_\odot$ yr$^{-1}$, \citet{Williams1987}) and large-amplitude IR variables (e.g. \citet{Williams1990}). Dust production in these systems is thought to originate from the compression of stellar winds in shocked wind-collision regions between the WC star and a luminous O or B companion \citep{Usov1991}. Some dusty WC systems therefore exhibit periodic or episodic IR brightening events corresponding to dust production. For example, the archetypal periodic dust-making Wolf-Rayet binary WR140 (WC7+O5I), located at $d =  1.67 \pm0.03$ kpc \citep{Monnier2011}, has an eccentric orbit and exhibits IR outbursts during periastron passage in its 7.94 yr orbit \citep{Williams2009}. WR140 has mean infrared magnitudes of L = 3.5 and M = 3.3 and an amplitude $\Delta M = 2.4$ (\citet{Williams2009,Taranova2011}). On Fig. \ref{fig:pldiagram} this would put it among the very longest period variables with $M_{[4.5]}\sim -8$, but there are more IR luminous WR binaries. The variable, dusty ``pinwheel" system WR 98a (WC8-9) \citep{Monnier1999} is located at $d = 1.9$ kpc and exhibits a mean infrared magnitude of L = 1.5 ($M_{[3.6]} \sim-10$) with a 1.6 yr orbital period \citep{Williams1995}. One of the most IR luminous WR binaries, WR 48a (WC8 + O), shows periodic variability on 32 yr timescales with peak L$^{\prime}$ and M magnitudes of 0.1 and $-0.7$, respectively \citep{Williams2012}. At a distance of 4 kpc, the absolute IR magnitude of WR48a is $M_{[4.5]} \sim -14$, which is consistent with some of the brightest variables in Fig. \ref{fig:pldiagram}.

Some WR binary systems exhibit a different type of variability : deep optical eclipses that do not phase with their orbital period \citep{Williams2014}. Unlike the periodic variability linked to enhanced dust formation at periastron, the erratic eclipses are likely due to obscuration by dust formed in clumps along the line of sight. WR104, a WC9+O binary surrounded by the Pinwheel Nebula \citep{Tuthill1999}, has $[3.6] = -0.13$ and $[4.8] = -1.01$ \citep{Gehrz1974,Williams1987} and exhibits large amplitude and erratic optical variability \citep{Williams2014}. However, there is only limited information on its infrared variability. Using a distance of $2.58 \pm 0.12$ kpc \citep{Soulain2018} we estimate $M_{[4.5]} \sim -13$, i.e., comparable to the brightest variables under discussion (the Gaia DR2 parallax, which has a large uncertainty, would make it more distant and hence even brighter).

It is not easy to identify stars in the common envelope phase that must be the penultimate stage in the evolution of many types of close binary \citep{Han1995}. One-dimensional hydrodynamic simulations \citep{Clayton2017} of low mass red giants undergoing a common envelope event indicate that their envelopes become unstable and develop large amplitude pulsations; these may be quasi-periodic and can drive high mass-loss rates. \citet{Glanz2018} also discuss the similarity of common envelope stars to pulsating AGB stars, how dust could form in their extended atmospheres and drive high mass-loss rates.
The TZOs are postulated examples of common envelope evolution. These would be neutron stars embedded in an extended hydrogen envelope that form from a massive binary after one star explodes as a supernova and its remnant core becomes embedded in a common envelope with its companion \citep{Thorne1975}. It has also been suggested that LBVs could be TZOs \citep{King2000}.

Fig. \ref{fig:pldiagram} shows the SMC source MSX SMC055 (IRAS 00483$-$7347), which is both a super-AGB candidate \citep{Groenewegen2009} and a TZO candidate \citep{Levesque+2014}. Deriving the initial mass of the star from the available data is not trivial and it is therefore not possible to be certain which of these interpretations is correct. MSX SMC055 has a period of 1859 days \citep{Soszynski2011} and an $L$ amplitude of 0.7  \citep{Menzies2019} and in Fig. 2 it falls very close to the clump of SPIRITS variables that seem to be on an extension of the AGB fundamental sequence. We therefore suggest that these may be super-AGB stars (and/or TZOs).






\begin{table*}
\begin{threeparttable}
\caption{Variable Catalog}
\label{tab:variable_catalog}
\begin{tabular}{lllllccllll}
\toprule
Name & RA & Dec & Gal & Period & $M_{[3.6]}$ & $M_{[4.5]}$ & $\Delta$[3.6] & $\Delta$[4.5] & Quality & Ref~* \\
\midrule
14apu & 210.882379 & 54.373631 & M101 & 1752  & -14.93 +/- 0.06 & -15.51 +/- 0.03 & 0.29  & 0.25 & silver & 1\\
14aue & 56.70243  & 68.067607 & IC342 & 344  & -14.31 +/- 0.03 & -14.80 +/- 0.03 & 0.14  & 0.11 & gold & 2\\
15ty & 308.730675 & 60.135467 & N6946 & 1587  & -13.97 +/- 0.04 & -14.80 +/- 0.06 & 0.47  & 0.43 & gold & 2, 3, 4\\
15pk & 210.885626 & 54.295308 & M101 & 1750  & -13.38 +/- 0.03 & -14.20 +/- 0.07 & 0.24  & 0.29 & bronze & 1\\
18ec & 202.512772 & 47.170201 & M51a & 2538  & -13.42 +/- 0.08 & -13.97 +/- 0.07 & 0.54  & 0.60 & gold & 5\\
14axv & 287.472939 & -63.823283 & N6744 & 381  & -12.82 +/- 0.07 & -13.62 +/- 0.08 & 1.75  & 1.88 & bronze & SN2005at\\
15js & 204.211515 & -29.873467 & M83 & 1603  & -12.87 +/- 0.03 & -13.59 +/- 0.03 & 0.95  & 0.86 & silver & 6\\
14apz & 210.718431 & 54.336889 & M101 & 1397  & -12.00 +/- 0.12 & -13.48 +/- 0.04 & 1.52  & 0.99 & gold & 1, 2\\
16do & 204.279883 & -29.89054 & M83 & 2071  & -12.35 +/- 0.07 & -13.47 +/- 0.04 & 0.36  & 0.42 & silver & \\
15jt & 204.284774 & -29.848441 & M83 & 928  & -13.00 +/- 0.11 & -13.42 +/- 0.19 & 0.53  & 0.76 & silver & 8\\
16po & 204.256266 & -29.907444 & M83 & 1206  & -12.56 +/- 0.03 & -13.35 +/- 0.05 & 0.39  & 0.32 & bronze & 9\\
14aq & 149.082516 & 69.06756  & M81 & 824  & -12.17 +/- 0.02 & -12.97 +/- 0.03 & 0.61  & 0.23 & bronze & \\
14apy & 210.718747 & 54.297112 & M101 & 1679  & -12.21 +/- 0.06 & -12.94 +/- 0.09 & 1.16  & 1.56 & silver & 1\\
14aqa & 210.803966 & 54.362223 & M101 & 1682  & -11.16 +/- 0.11 & -12.94 +/- 0.05 & 1.56  & 0.92 & bronze & 1, 10\\
15agi & 56.505276 & 68.092698 & IC342 & 929  & -12.09 +/- 0.07 & -12.93 +/- 0.07 & 0.45  & 0.51 & gold & \\
14bpf & 204.27225 & -29.883631 & M83 & 1234  & -11.92 +/- 0.04 & -12.89 +/- 0.08 & 0.82  & 0.88 & gold & \\
16pn & 204.292494 & -29.879432 & M83 & 1158  & -11.91 +/- 0.04 & -12.84 +/- 0.04 & 0.60   & 0.64 & silver & \\
15zg & 204.269869 & -29.843227 & M83 & 1127  & -11.87 +/- 0.05 & -12.77 +/- 0.04 & 0.63  & 0.79 & gold & 7, 9, 11\\
15ke & 204.258726 & -29.879923 & M83 & 857  & -12.86 +/- 0.06 & -12.73 +/- 0.04 & 0.29  & 0.45 & gold & 8, 12\\
14bot & 204.170951 & -29.874657 & M83 & 1666  & -12.60 +/- 0.11 & -12.65 +/- 0.04 & 0.44  & 1.11 & gold & \\
15kz & 210.954843 & 54.369441 & M101 & 1496  & -11.55 +/- 0.07 & -12.62 +/- 0.06 & 1.29  & 1.22 & silver & 1\\
15oe & 287.456721 & -63.819975 & N6744 & 2372  & -11.17 +/- 0.10 & -12.58 +/- 0.04 & 3.40   & 2.47  & silver & \\
14aul & 56.533417 & 68.105675 & IC 342 & 997  & -11.90 +/- 0.04 & -12.57 +/- 0.05 & 0.62  & 0.68  & gold & \\
15jw & 204.221199 & -29.875165 & M83 & 1787  & -12.06 +/- 0.07 & -12.57 +/- 0.07 & 0.65  & 1.05 & gold & 8, 12, 13\\
14akr & 204.224543 & -29.864975 & M83 & 1964  & -11.73 +/- 0.06 & -12.54 +/- 0.07 & 1.22  & 1.06 & silver & \\
15ky & 210.75046 & 54.381013 & M101 & 897  & -12.35 +/- 0.04 & -12.51 +/- 0.06 & 0.53  & 0.83 & gold & 1\\
16dp & 204.24353 & -29.896648 & M83 & 1215  & -12.06 +/- 0.06 & -12.50 +/- 0.08 & 0.64  & 0.81 & gold & \\
14bco & 210.727875 & 54.288666 & M101 & 1411  & -11.71 +/- 0.03 & -12.50 +/- 0.04 & 1.63  & 1.08 & silver & \\
14akl & 204.183788 & -29.823974 & M83 & 1506  & -11.88 +/- 0.01 & -12.49 +/- 0.02 & 0.88  & 0.80 & silver & \\
15kd & 204.235358 & -29.871858 & M83 & 1033  & -11.78 +/- 0.06 & -12.48 +/- 0.06 & 1.26  & 1.21 & silver & \\
14ajs & 204.311239 & -29.834511 & M83 & 1472  & -11.25 +/- 0.07 & -12.46 +/- 0.05 & 2.46  & 1.35 & silver & \\
18ae & 210.824216 & 54.314684 & M101 & 1530  & -11.24 +/- 0.07 & -12.45 +/- 0.05 & 5.23  & 1.78 & gold & 2\\
15kc & 204.20733 & -29.838336 & M83 & 1065  & -11.99 +/- 0.03 & -12.43 +/- 0.03 & 0.75  & 0.45 & gold & 12, 13\\
14bqb & 204.253607 & -29.917184 & M83 & 1673  & -11.67 +/- 0.07 & -12.42 +/- 0.04 & 1.54  & 1.30 & bronze & 6, 9, 14\\
14apq & 148.775969 & 69.127891 & M81 & 1606  & -11.68 +/- 0.05 & -12.40 +/- 0.02 & 1.96  & 1.60 & silver & \\
14atg & 204.204003 & -29.887841 & M83 & 772  & -12.00 +/- 0.03 & -12.40 +/- 0.03 & 0.86  & 1.20 & silver & 15\\
15zh & 204.296821 & -29.863899 & M83 & 1731  & -11.27 +/- 0.03 & -12.33 +/- 0.02 & 3.17  & 2.62 & silver & \\
17fo & 184.709724 & 47.303198 & M106 & 1542  & -11.16 +/- 0.09 & -12.30 +/- 0.04 & 2.04  & 1.67 & gold & \\
14bov & 204.296388 & -29.867147 & M83 & 1656  & -11.27 +/- 0.03 & -12.29 +/- 0.02 & 1.08  & 1.29 & silver & \\
14boy & 204.246034 & -29.811423 & M83 & 1440  & -11.06 +/- 0.08 & -12.29 +/- 0.09 & 1.37  & 0.87 & silver & \\
15km & 204.241521 & -29.843867 & M83 & 1362  & -11.80 +/- 0.06 & -12.28 +/- 0.08 & 0.68  & 0.77 & gold & \\
14bcp & 210.748934 & 54.295433 & M101 & 1592  & -11.61 +/- 0.05 & -12.27 +/- 0.08 & 0.85  & 1.32 & gold & 1\\
17kh & 204.298281 & -29.828349 & M83 & 1218  & -11.53 +/- 0.08 & -12.25 +/- 0.08 & 0.79  & 0.58 & gold & \\
14akp & 204.274805 & -29.94043 & M83 & 1507  & -11.41 +/- 0.06 & -12.23 +/- 0.03 & 1.44  & 1.48 & silver & \\
14ck & 148.942537 & 68.995879 & M81 & 1610  & -11.13 +/- 0.06 & -12.22 +/- 0.04 & 2.00    & 1.53 & gold & \\
14oj & 148.726763 & 69.068093 & M81 & 2956  & -10.82 +/- 0.06 & -12.22 +/- 0.06 & 0.50   & 0.62  & bronze & 2\\
14akn & 204.270461 & -29.908382 & M83 & 1430  & -11.53 +/- 0.05 & -12.16 +/- 0.04 & 1.02  & 1.20  & silver & 4, 9, 14\\
16dt & 204.275987 & -29.88318 & M83 & 1089  & -11.08 +/- 0.05 & -12.16 +/- 0.03 & 1.08  & 0.80  & gold & 16\\
14ajf & 210.844714 & 54.282388 & M101 & 1527  & -11.27 +/- 0.07 & -12.16 +/- 0.12 & 1.22  & 1.42 & gold & 1\\
16ds & 204.254718 & -29.912643 & M83 & 1377  & -11.43 +/- 0.05 & -12.10 +/- 0.04 & 0.72  & 0.94  & silver & 9, 14\\
15mn & 210.955989 & 54.347408 & M101 & 1661  & -10.66 +/- 0.04 & -12.10 +/- 0.03 & 2.33  & 1.48 & silver & 1\\
15kj & 204.215737 & -29.849034 & M83 & 1450  & -11.64 +/- 0.04 & -12.08 +/- 0.04 & 0.84  & 1.30  & silver & \\
15ki & 204.189008 & -29.85111 & M83 & 1119  & -11.41 +/- 0.04 & -12.05 +/- 0.03 & 1.06  & 0.92  & gold & 2, 13\\
14aks & 204.206609 & -29.79767 & M83 & 1426  & -11.14 +/- 0.04 & -12.04 +/- 0.07 & 1.33  & 1.26  & silver &\\
15zi & 204.267489 & -29.802396 & M83 & 1771  & -10.90 +/- 0.06 & -12.04 +/- 0.08 & 2.49  & 2.03  & silver &\\
16dr & 204.303412 & -29.85154 & M83 & 1509  & -11.05 +/- 0.07 & -12.04 +/- 0.05 & 1.52  & 1.66  & silver & 2\\
14ni & 148.873272 & 69.114964 & M81 & 1408  & -11.05 +/- 0.09 & -12.01 +/- 0.11 & 2.44  & 1.19 & bronze & 17\\
15kb & 204.186924 & -29.863754 & M83 & 1458 & -11.34 +/- 0.03 & -12 +/- 0.02 & 1.17 & 1.28 & gold & \\
...\\
\bottomrule
\end{tabular}
\begin{tablenotes}
\small
\item Description of columns :\\
$Name :$ SPIRITS name for variable; 
$RA$ and $Dec :$ RA and Dec (in degrees for equinox 2000);
$Gal :$ Galaxy Name;
$Period :$ Period in days; \\
$[3.6] :$ 3.6$\mu$m mean absolute magnitude; 
$[4.5] :$ 4.5$\mu$m mean absolute magnitude; 
$\Delta[3.6] :$ 3.6$\mu$m amplitude (mags); 
$\Delta[4.5] :$ 4.5$\mu$m amplitude (mags);
\emph{Quality} : Subjective classification for variable;
\emph{Ref} : Associations with clusters or other catalogs.\\
* List of references in \emph{Ref}:\\
$[1] : $ \citet{Grammer2013}, $[2] : $ \citet{Page2014}, $[3] : $ \citet{Donovan_Meyer2013}, $[4] : $ \citet{Larsen2004}, $[5] : $ \citet{Lee2011}, $[6] : $ \citet{Ryon2015}, $[7] : $ \citet{Flesch2017}, $[8] : $ \citet{Vucetic2015}, $[9] : $ \citet{Silva-Villa2011}, $[10] : $ \citet{Drazinos2013}, $[11] : $ \citet{Larsen2011}, $[12] : $ \citet{Blair2012}, $[13] : $ \citet{Hadfield2005}, $[14] : $ \citet{Mora2009}, $[15] : $ \citet{Larsen1999}, $[16] : $ \citet{HEASARC2004}, $[17] : $ \citet{Khan2010}.\\ 
The complete catalog will be available as a machine readable table.
\end{tablenotes}
\end{threeparttable}
\end{table*}

\section{Summary}
We present a catalog of 417 luminous IR variable candidates in 20 nearby luminous galaxies targeted by the ongoing SPIRITS survey. We report the periods, mean magnitudes and the reference photometry for these variables. We also present the IRAC 3.6, 4.5, 5.8 and 8.0 $\mu$m PSF catalogs for reference images of all 20 host galaxies. Over 300 of these are 
AGB candidates, based on their positions in the period-luminosity and color-magnitude diagrams. 

In addition to these, we find about 50 variables that are more luminous and have longer periods than those from previous surveys. The majority of these have $1000<P<2000$ and $M_{[4.5]}$ between $-11$ and $-13$, and fall near the extrapolation of the relation for fundamental pulsation in the period-luminosity diagram. This suggests that these are AGB stars with  massive progenitors, certainly more massive than those of the LMC OH/IR stars, and perhaps even super-AGBs. The sample with M$_{[4.5]}$ brighter than $-12$ will include some RSGs with high mass-loss rates and probably exotic examples of interacting binaries, e.g., TZOs. There are also nine long-period variables with M$_{[4.5]}$ brighter than $-13$, that lie in a previously unexplored region of the period-luminosity diagram; several of these are located within young clusters.   
The extra-luminous variables will include some contamination that is inevitable given our limited spatial resolution, but some of these sources could also be part of a new, previously unknown class of variables. Exploring the true nature of these extra-luminous variables will be an exciting avenue of exploration for future missions, such as the James Webb Space Telescope.  


\section*{Acknowledgements}
We thank Dave Cook, Maria Drout, Shazrene Mohamed and John Menzies for useful discussions. This work is based on observations made with \emph{Spitzer}, which is operated by the Jet Propulsion Laboratory, California Institute of Technology under a contract with NASA. The SPIRITS team acknowledges generous support from the NASA \emph{Spitzer} grants for SPIRITS. PAW thanks the South African NRF for research funding. JEJ acknowledges support from the National Science Foundation Graduate Research Fellowship under Grant No.\ DGE-1144469. RDG was supported by NASA and the United States Air Force. This research has made use of the VizieR catalogue access tool, CDS, Strasbourg, France. The original description of the VizieR service was published in A\& AS 143, 23. This work has made use of data from the European Space Agency (ESA) mission
{\it Gaia} (\url{https://www.cosmos.esa.int/gaia}), processed by the {\it Gaia}
Data Processing and Analysis Consortium (DPAC,
\url{https://www.cosmos.esa.int/web/gaia/dpac/consortium}). Funding for the DPAC
has been provided by national institutions, in particular the institutions
participating in the {\it Gaia} Multilateral Agreement.

\bibliography{references}
\bibliographystyle{mn2e}

\section*{Appendix A}
As described in \S \ref{sec:period_finding}, we assign a quality (gold, silver or bronze) to each of our bright variables based on visual inspection of light curves and periodograms. Here, we discuss some of the interesting variables in the bronze classification. 

\begin{figure}
	\includegraphics[width = 0.4\textwidth]{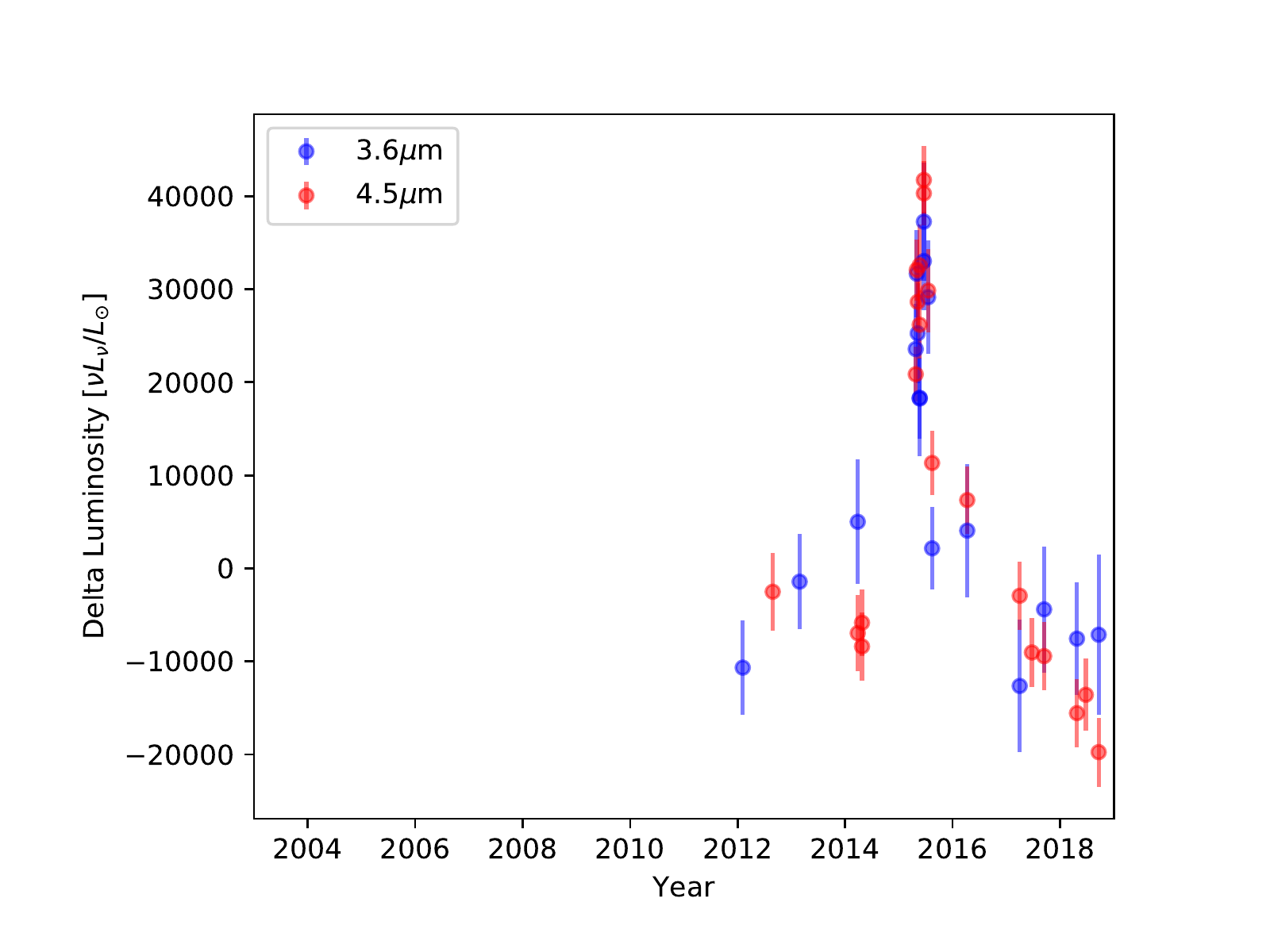}
    \caption{SPIRITS~15pk : M~101. This source is not a periodic variable, but seems to flare briefly.}
    \label{fig:15pk_lc}
\end{figure}

The source SPIRITS~15pk (Fig. \ref{fig:15pk_lc}) is not a periodic variable with period of 1750 days, but probably a brief flare (of $\approx$ 27 days). It is situated within a massive star cluster in M101 \citep{Grammer2013}. While low-mass YSOs are not prone to luminous outbursts, massive protostars can exhibit such flares and could be responsible for the brief outburst observed in this source. \citet{Bally2017} posit that a merger of a highly enshrouded protostar and a 15$M_{\odot}$ radio source could have powered a high energy ($10^{48}$ erg) explosion in the Orion OMC1 region about 500 years ago, producing a luminous infrared transient. Massive star mergers in star-forming regions could also result in high energy outbursts \citep{Bally2005}. Periodic flares are also possible in systems in which a massive companion has an eccentric, non-coplanar orbit with respect to a disk around the primary.

\begin{figure}
	\includegraphics[width = 0.4\textwidth]{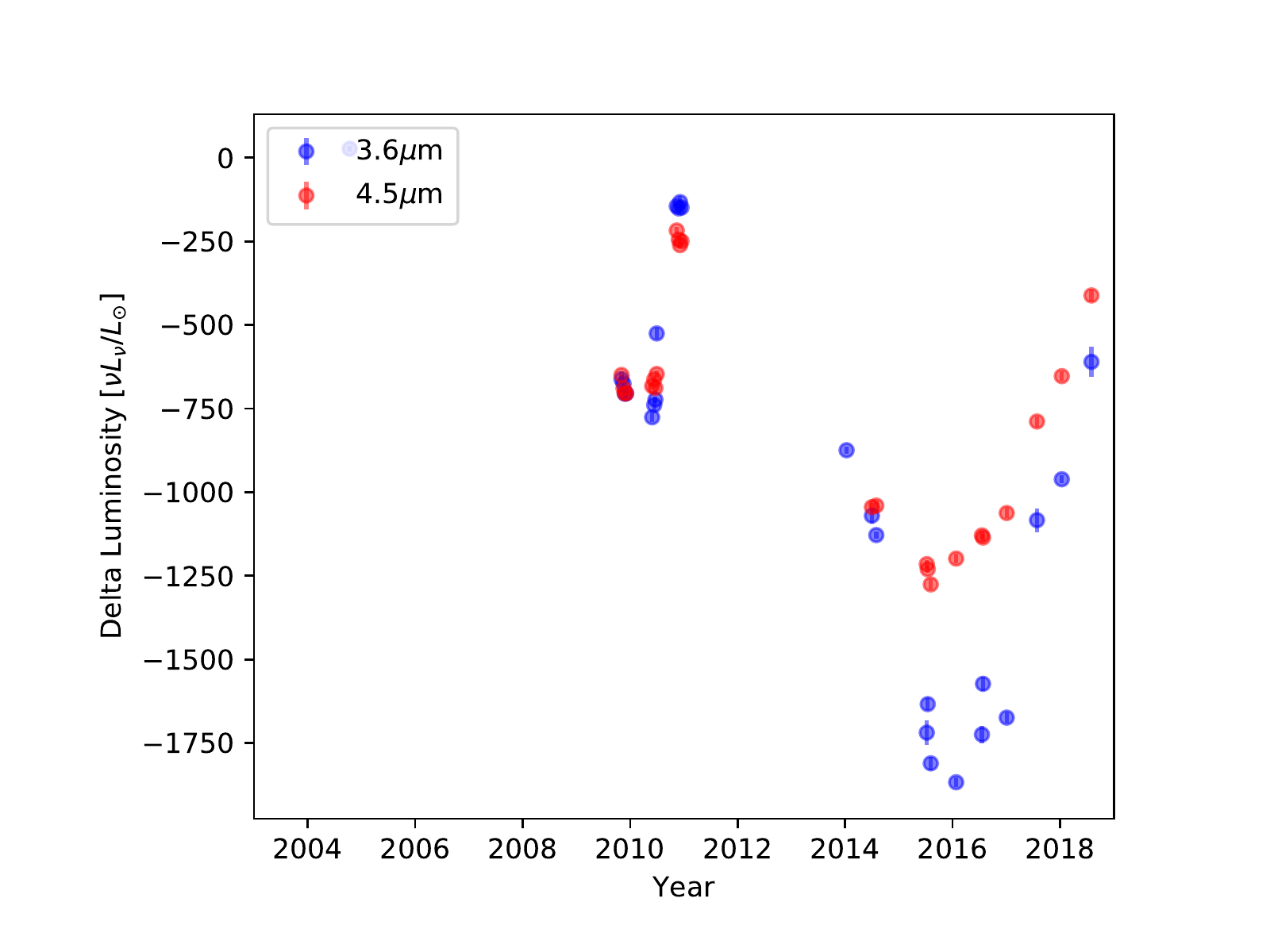}
    \caption{SPIRITS~14ct : NGC~6822. The light curve suggests shorter periods in addition to the 2947 days found in our analysis, see also Fig. \ref{fig:14ct}.}
    \label{fig:14ct_lc}
\end{figure}

\begin{figure}
	\includegraphics[width = 0.4\textwidth]{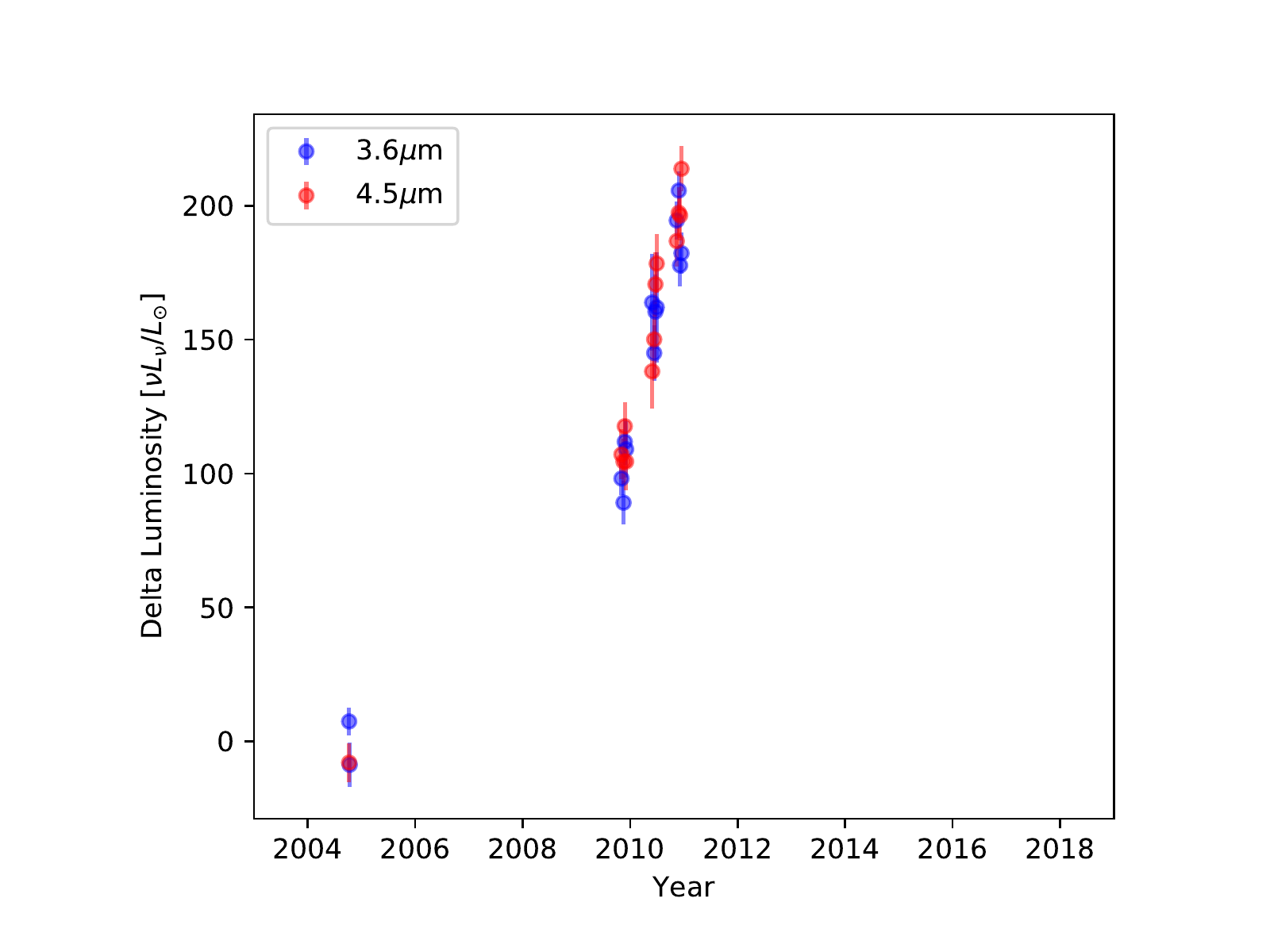}
    \caption{SPIRITS~14bkv : NGC~6822. This source has one epoch separated from the majority of the light curve, and this point influences the period we find. Many more observations are required to determine if this source is actually periodic.}
    \label{fig:14bkv_lc}
\end{figure}

\begin{figure}
	\includegraphics[width = 0.4\textwidth]{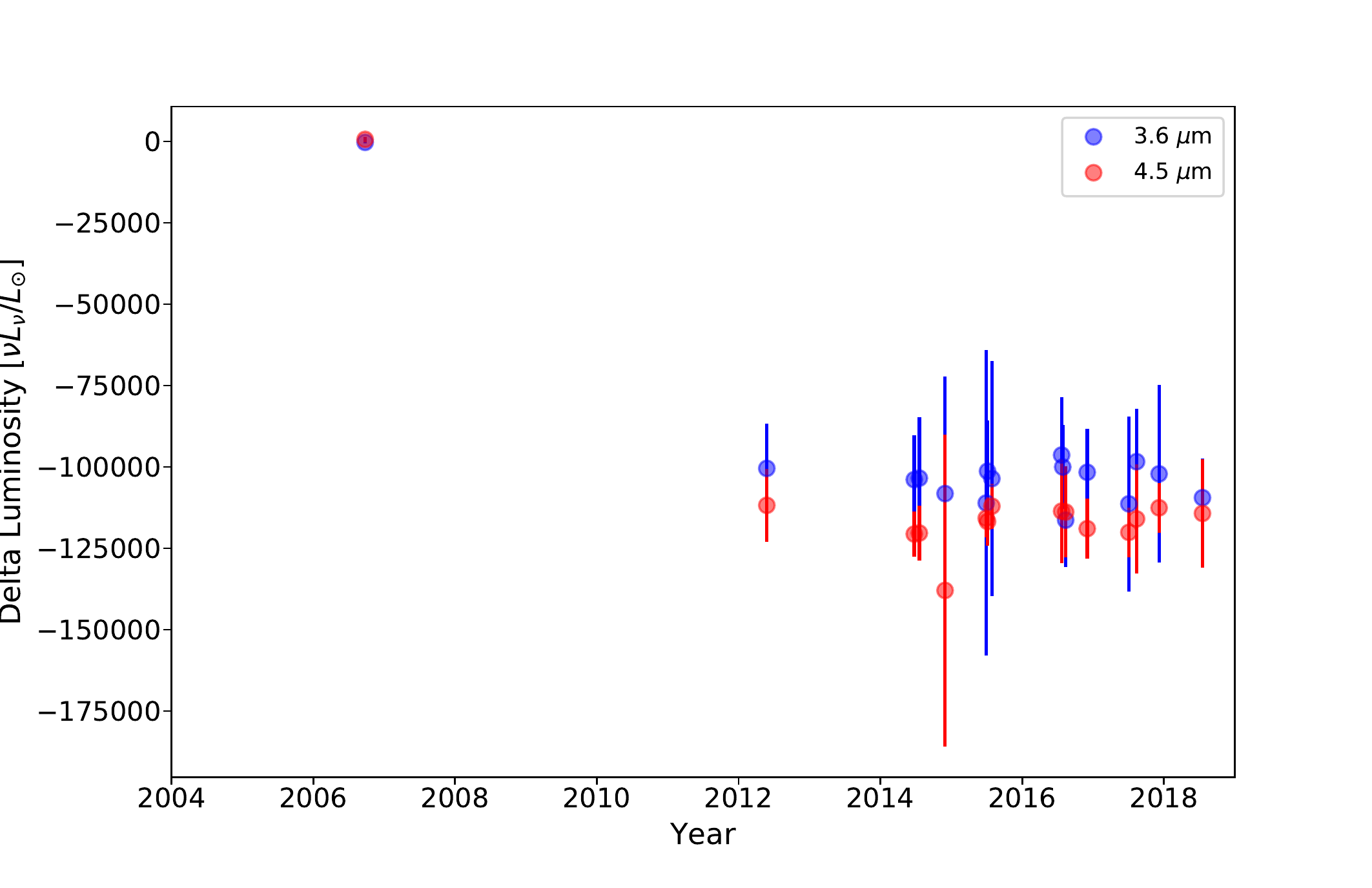}
    \caption{SPIRITS~14axv : NGC~6744. This source has one epoch from the reference image separated from the rest of the light curve, which drives the period finding algorithm. The source is SN2005at; it is obviously not periodic and only detected at the first epoch.}
    \label{fig:14axv_lc}
\end{figure}

\begin{figure}
	\includegraphics[width = 0.4\textwidth]{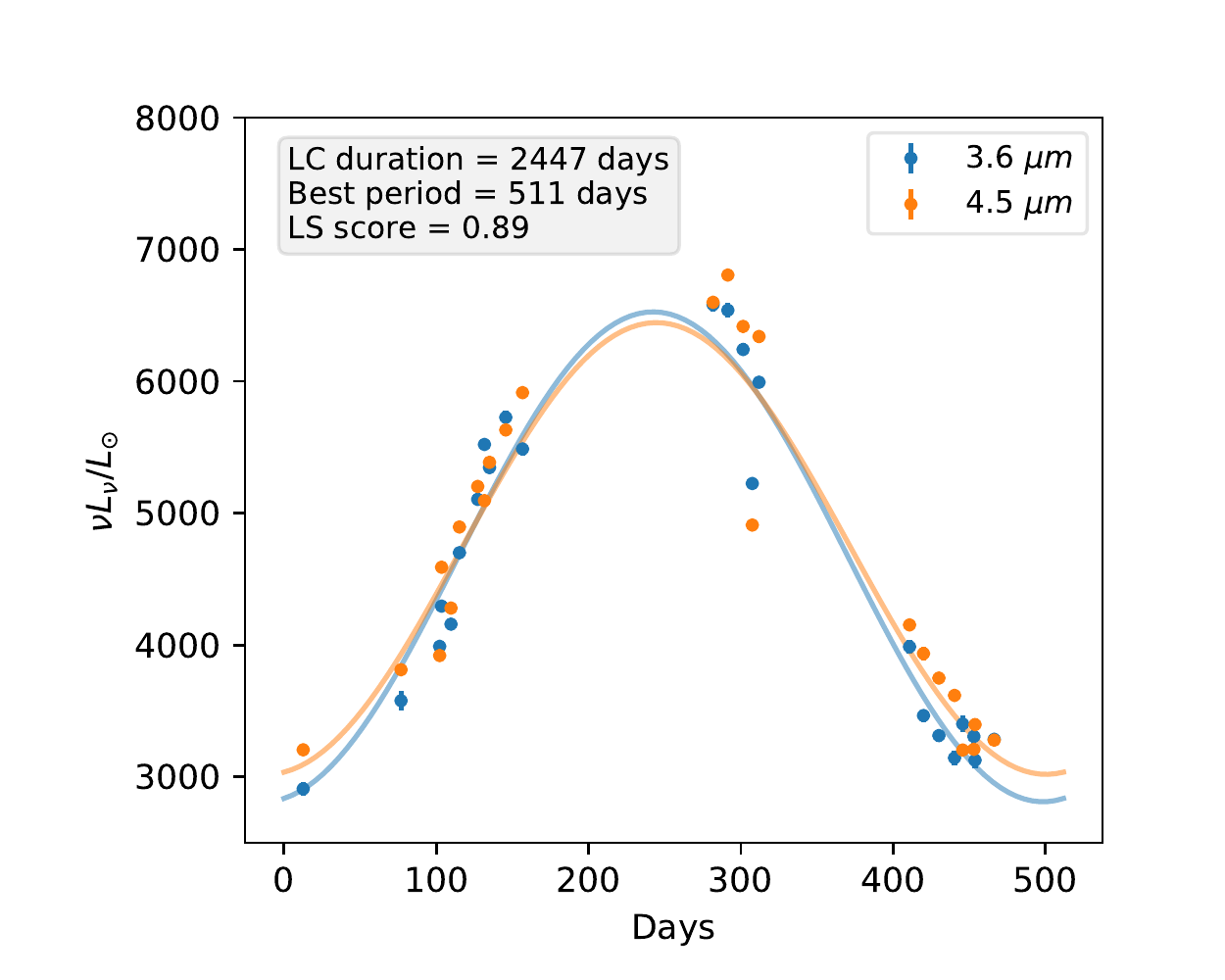}
    \caption{Phase$-$folded light curve of SPIRITS~14va : WLM. This source has a $M_{[4.5]} = -10.26$, $\Delta[3.6]=0.91$, $\Delta[4.5]=0.82$ and is a representative of the AGB candidates in our sample that appears in the same region as the LMC fundamental mode AGB variables in Fig. \ref{fig:pldiagram}.}
    \label{fig:14va}
\end{figure}

\begin{figure}
	\includegraphics[width = 0.4\textwidth]{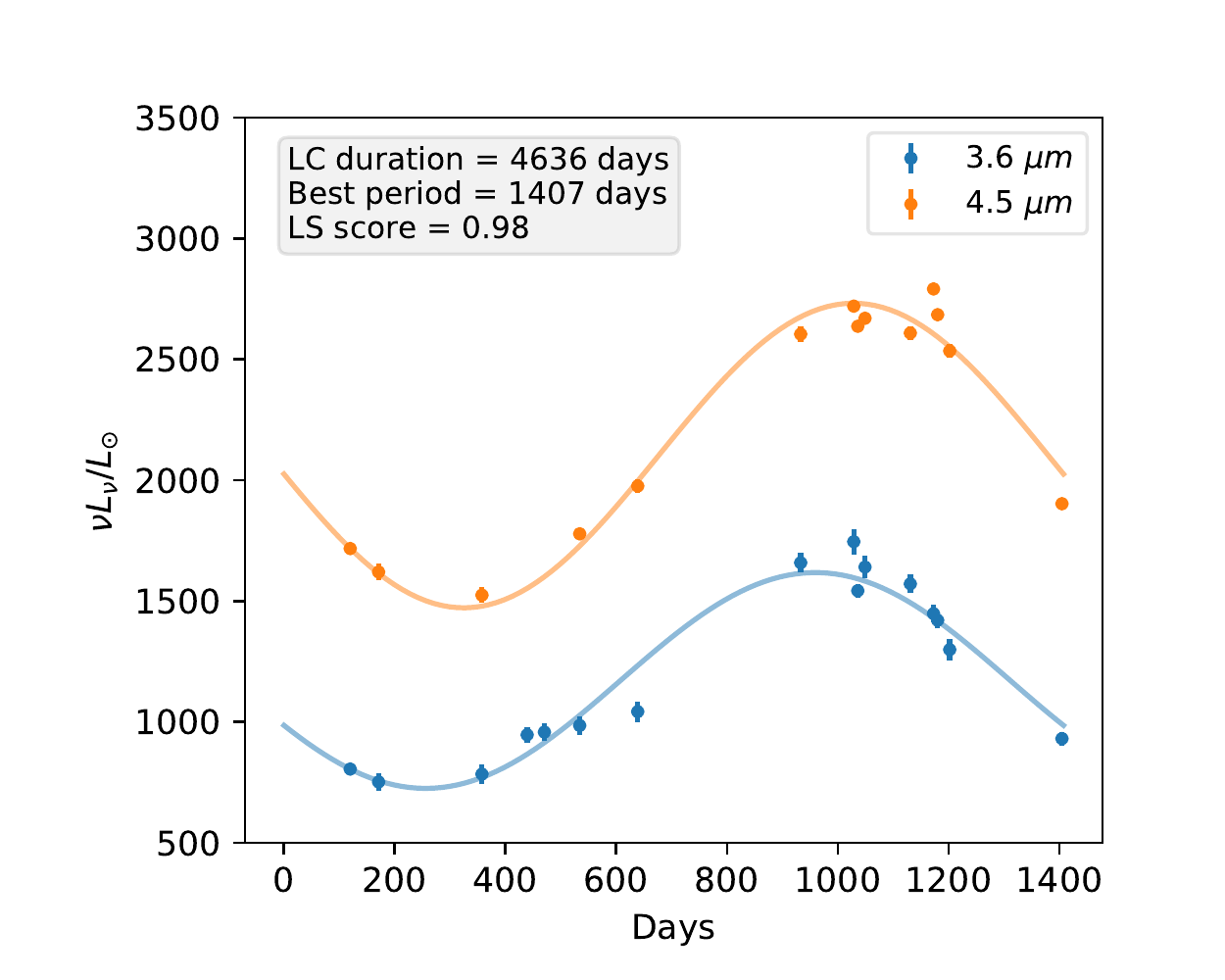}
    \caption{Phase$-$folded light curve of SPIRITS~15acg : NGC~185, one of the long period AGB candidates, that extends the long period sequence of LMC variables in Fig. \ref{fig:pldiagram} to higher periods. This source has $M_{[4.5]} = -9.38$, $\Delta[3.6]=0.87$ and $\Delta[4.5]=0.67$.}
    \label{fig:15acg}
\end{figure}

\begin{figure}
	\includegraphics[width = 0.4\textwidth]{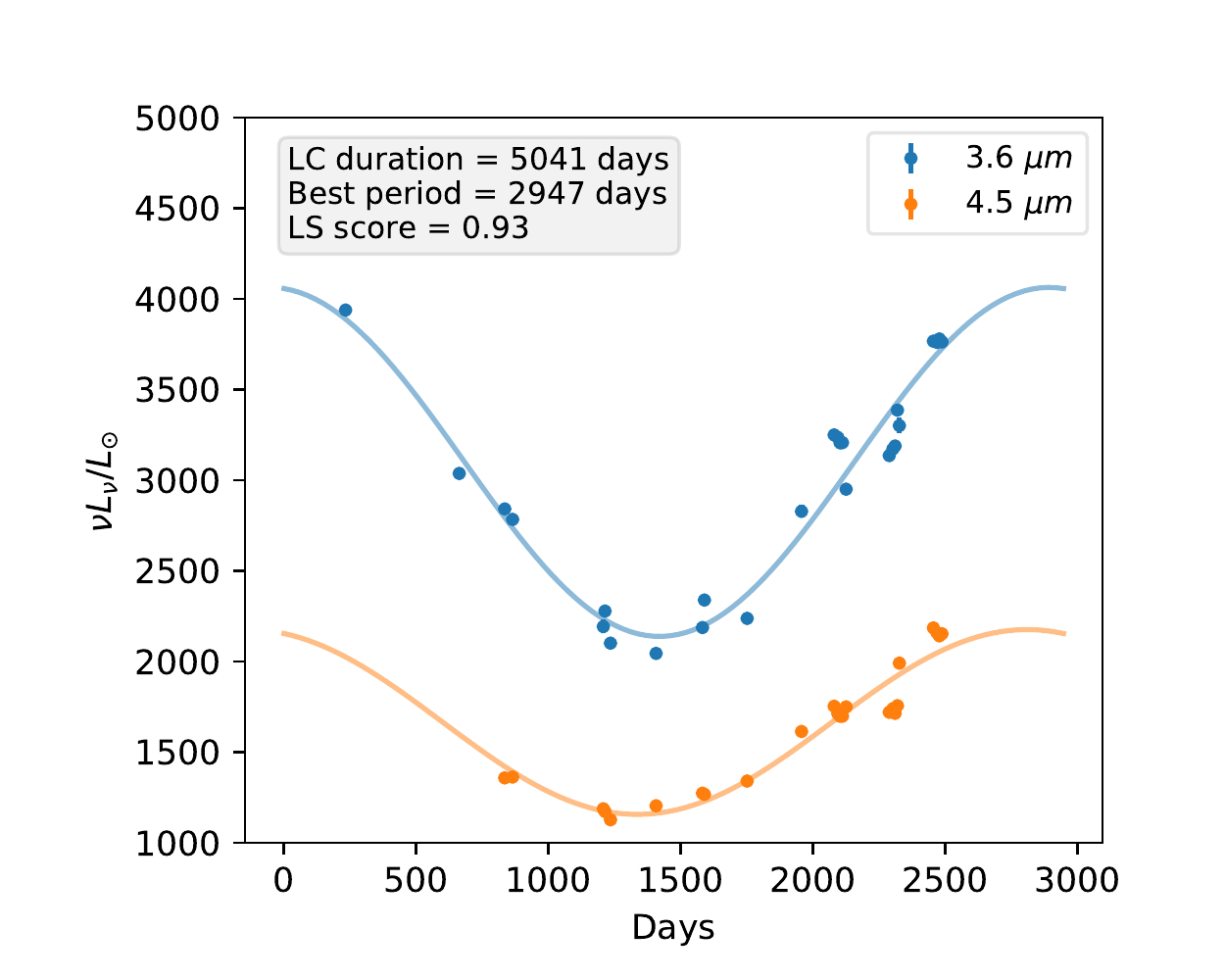}
    \caption{Phase$-$folded light curve of SPIRITS~14ct : NGC~6822, one of the long period AGB candidates; see also Fig. \ref{fig:14ct_lc}. This source has $M_{[4.5]} = -9.12$, $\Delta[3.6]=0.7$ and $\Delta[4.5]=0.69$}
    \label{fig:14ct}
\end{figure}

\begin{figure}
	\includegraphics[width = 0.4\textwidth]{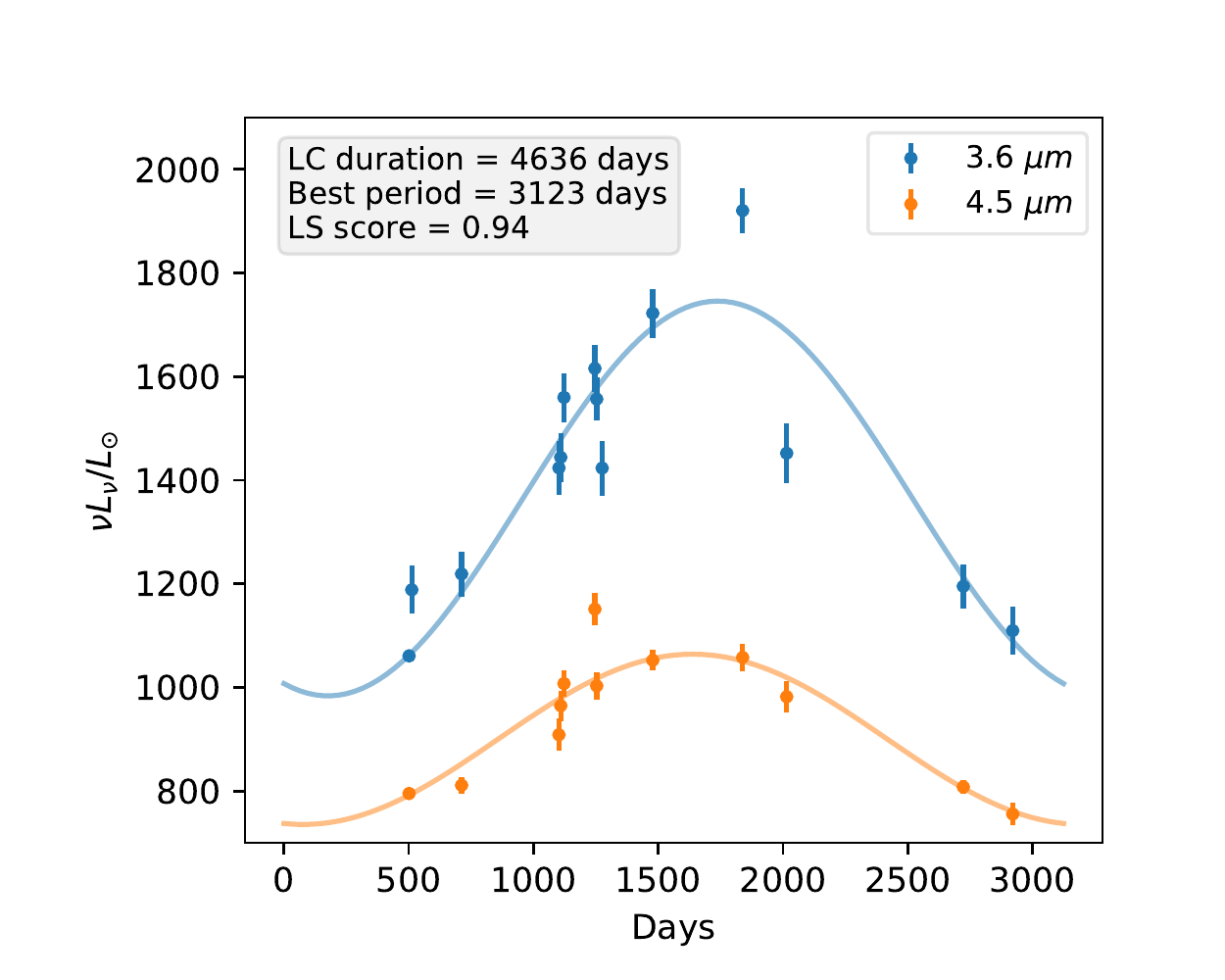}
    \caption{Phase$-$folded light curve of SPIRITS~15aci : NGC~185, the AGB candidate with longest period, that extends the long period sequence of LMC variables in Fig. \ref{fig:pldiagram} to longer periods. This source has $M_{[4.5]} = -8.46$, $\Delta[3.6]=0.62$ and $\Delta[4.5]=0.4$.}
    \label{fig:15aci}
\end{figure}

\begin{figure}
	\includegraphics[width = 0.4\textwidth]{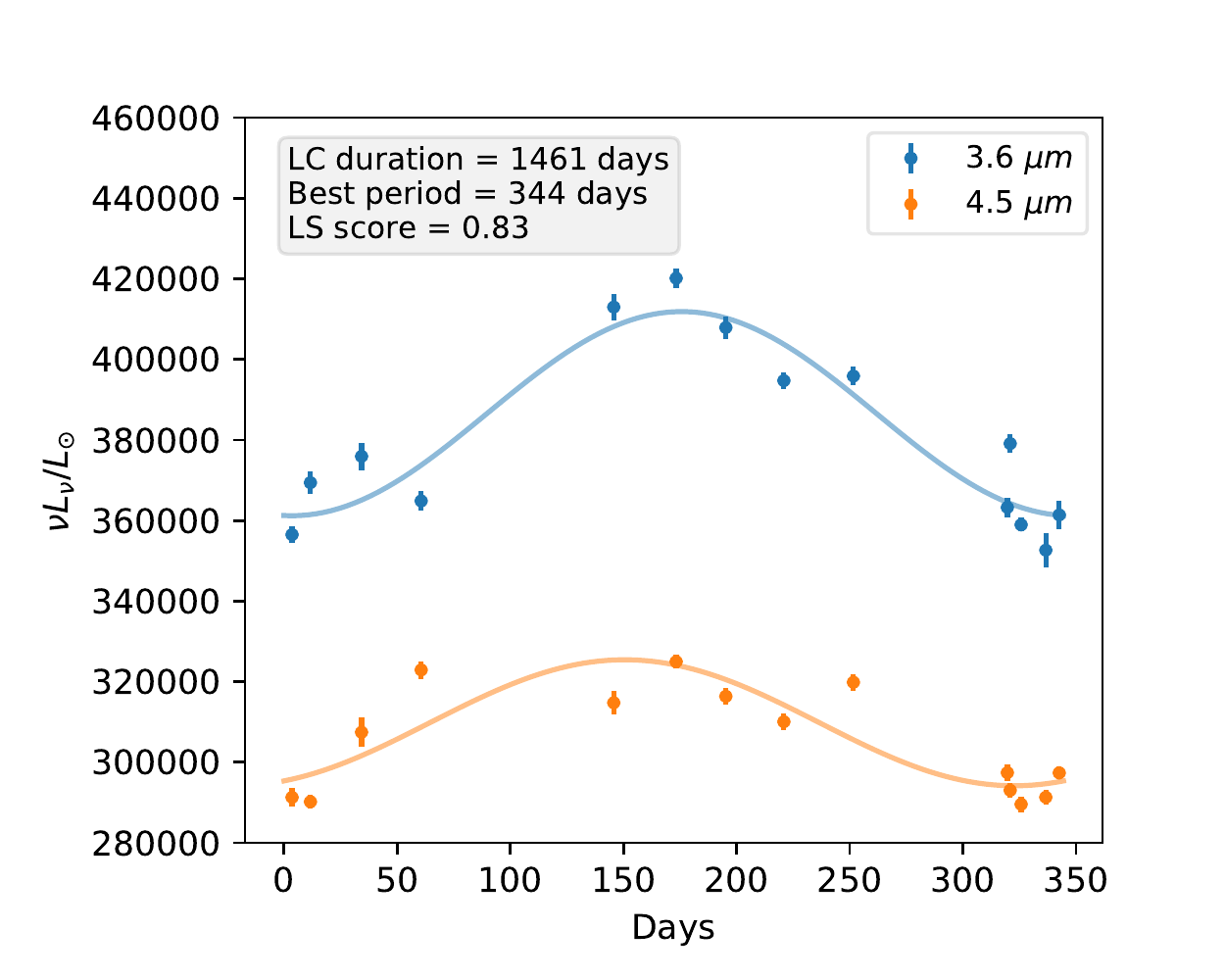}
    \caption{Phase$-$folded light curve of SPIRITS~14aue : IC342. This source is one of the two high luminosity ($M_{[4.5]}$ brighter than $-13$), low period variables in our sample, placed at the center of Fig. \ref{fig:pldiagram}. This soure has $M_{[4.5]} = -14.8$, $\Delta[3.6]=0.91$ and $\Delta[4.5]=0.11$.}
    \label{fig:14aue}
\end{figure}

\begin{figure}
	\includegraphics[width = 0.4\textwidth]{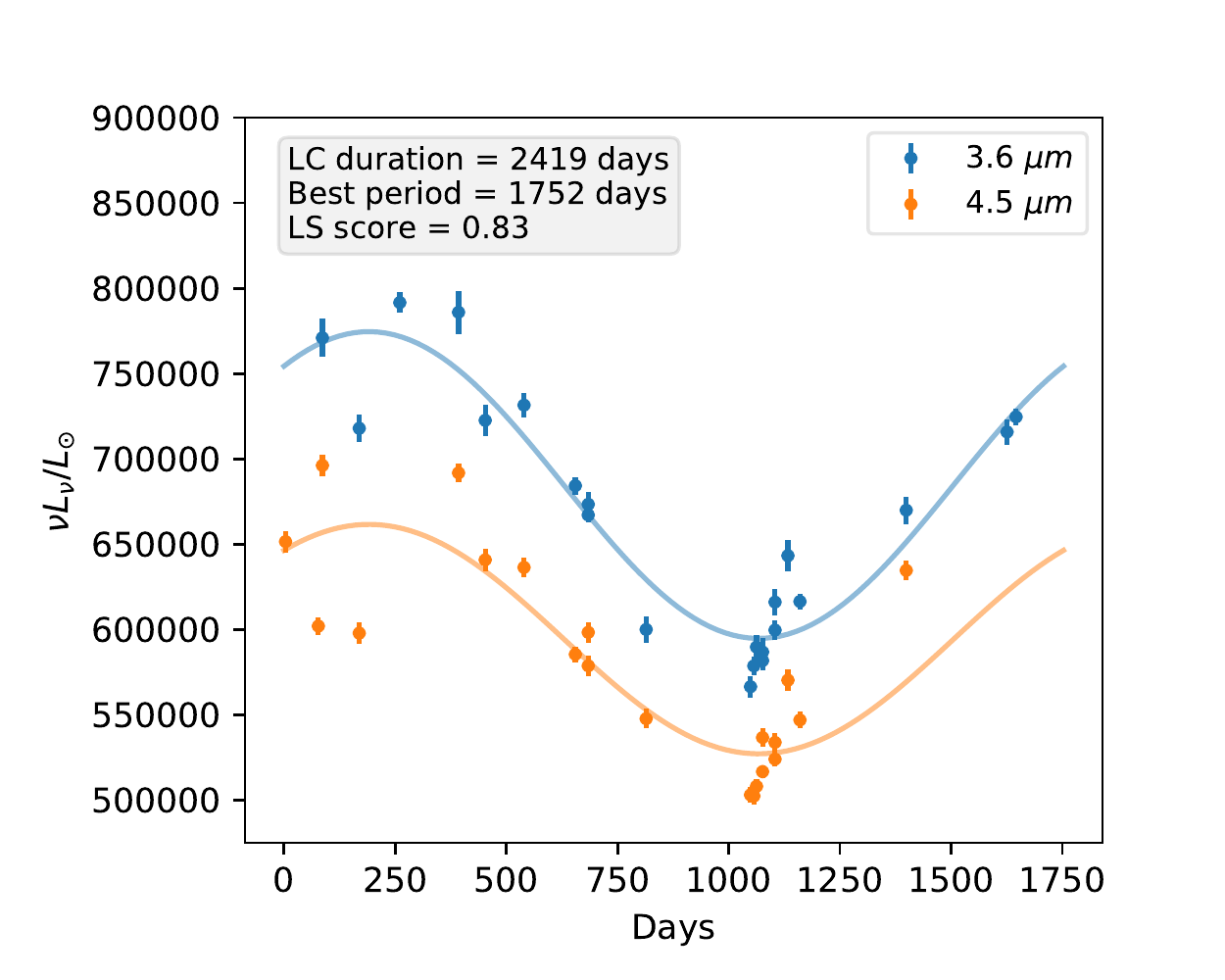}
    \caption{Phase$-$folded light curve of SPIRITS~14apu : M101, is the brightest source in our sample, with $M_{[4.5]} = -15.51$, $\Delta[3.6]=0.29$ and $\Delta[4.5]=0.25$. }
    \label{fig:14apu}
\end{figure}

\begin{figure}
	\includegraphics[width = 0.4\textwidth]{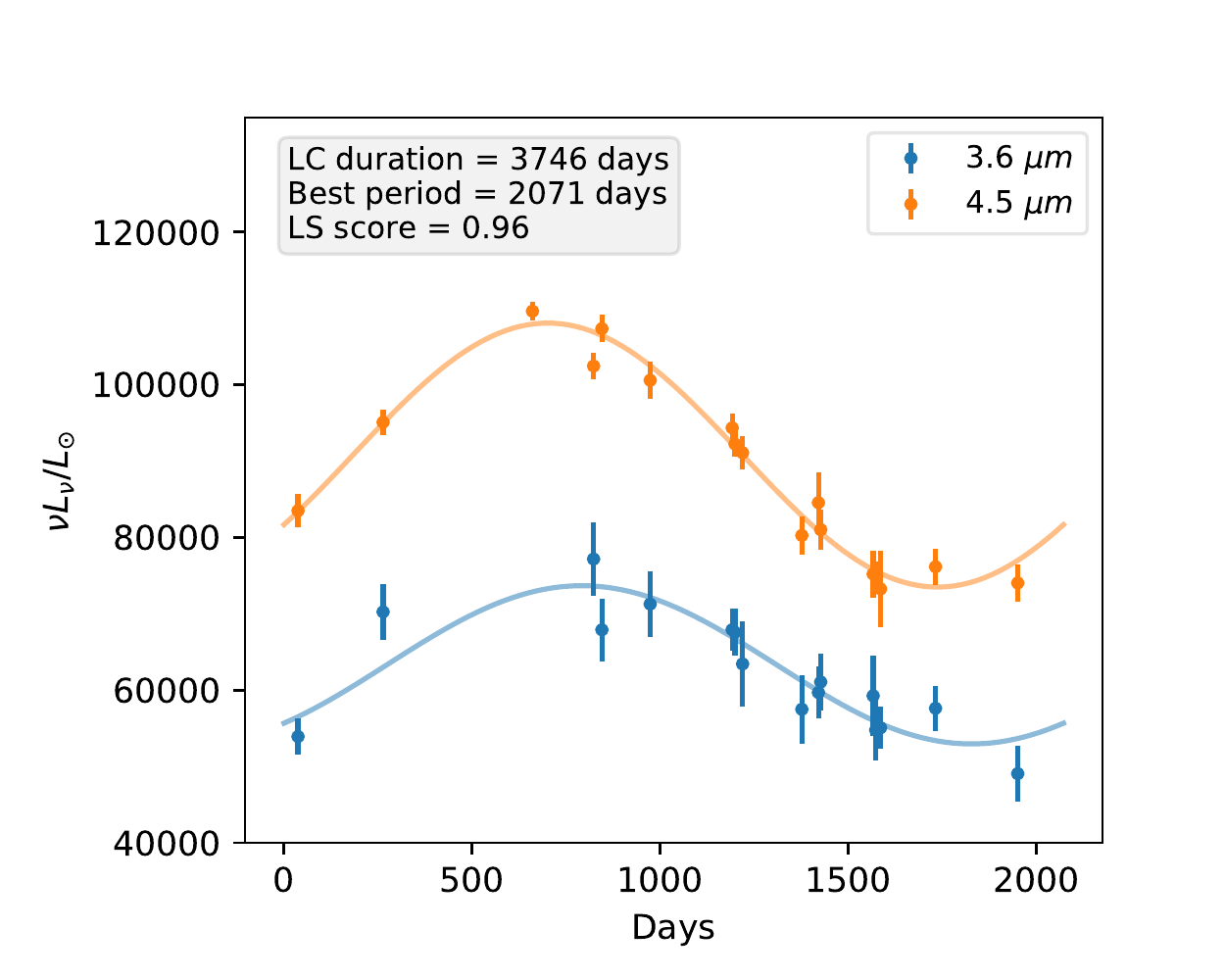}
    \caption{Phase$-$folded light curve of SPIRITS~16do : NGC~5236, one of the most luminous sources ($M_{[4.5]}$ brighter than $-13$) in our sample, with an extremely long period ($>2000$d). This source has $M_{[4.5]} = -13.47$, $\Delta[3.6]=0.36$ and $\Delta[4.5]=0.42$.}
    \label{fig:16do}
\end{figure}

\begin{figure}
	\includegraphics[width = 0.4\textwidth]{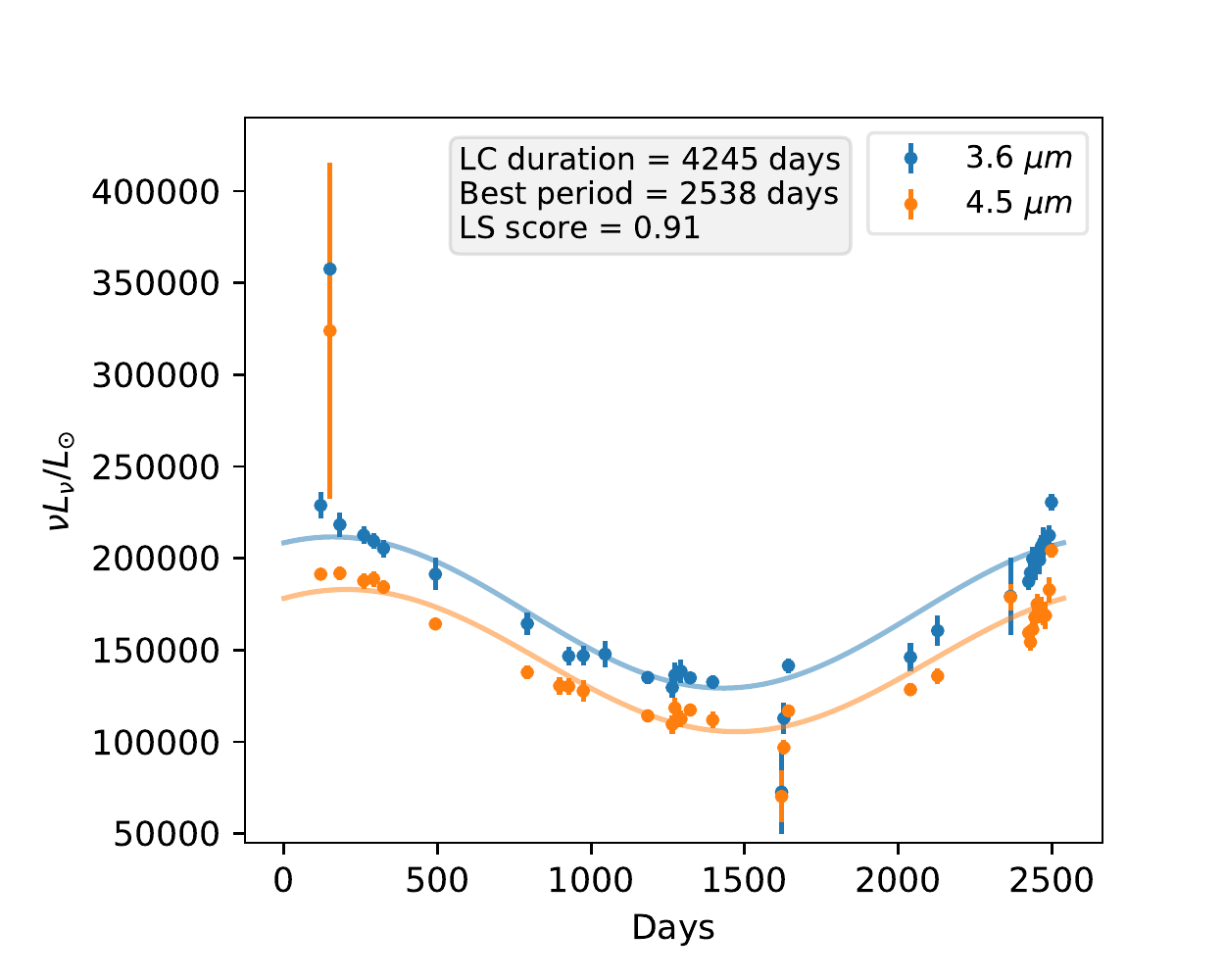}
    \caption{Phase$-$folded light curve of SPIRITS~18ec : NGC~5194, one of the most luminous sources ($M_{[4.5]}$ brighter than $-13$) in our sample, with an extremely long period ($>2500$d). This source has $M_{[4.5]} = -13.97$, $\Delta[3.6]=0.54$ and $\Delta[4.5]=0.6$.}
    \label{fig:18ec}
\end{figure}

\begin{figure}
	\includegraphics[width = 0.4\textwidth]{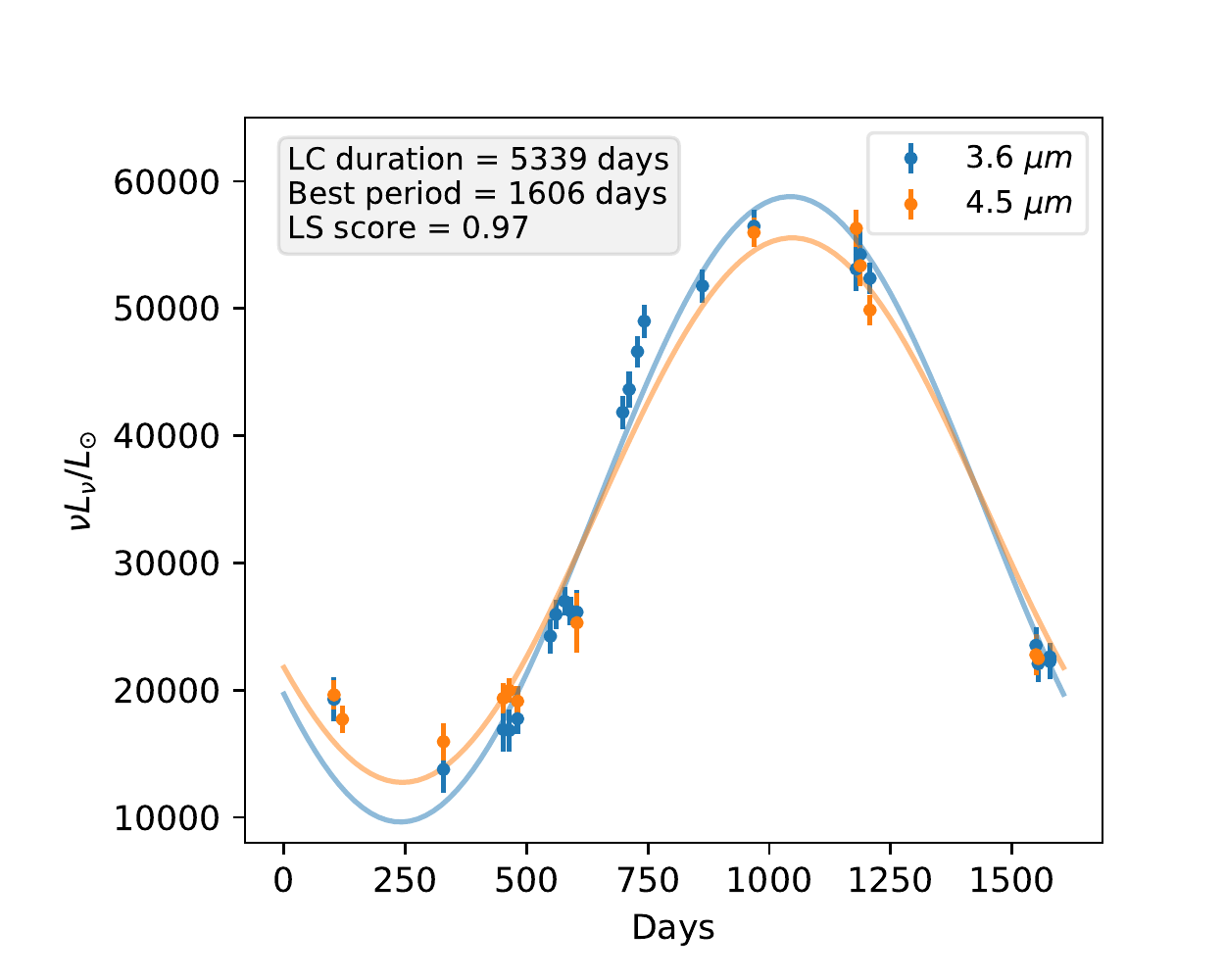}
    \caption{Phase$-$folded light curve of SPIRITS~14apq : M81, representative of sources with $M_{[4.5]}$ between $-12$ and $-13$. This source has $-12.4$, $\Delta[3.6]=1.96$ and $\Delta[4.5]=1.6$.}
    \label{fig:14apq}
\end{figure}

\begin{figure}
	\includegraphics[width = 0.4\textwidth]{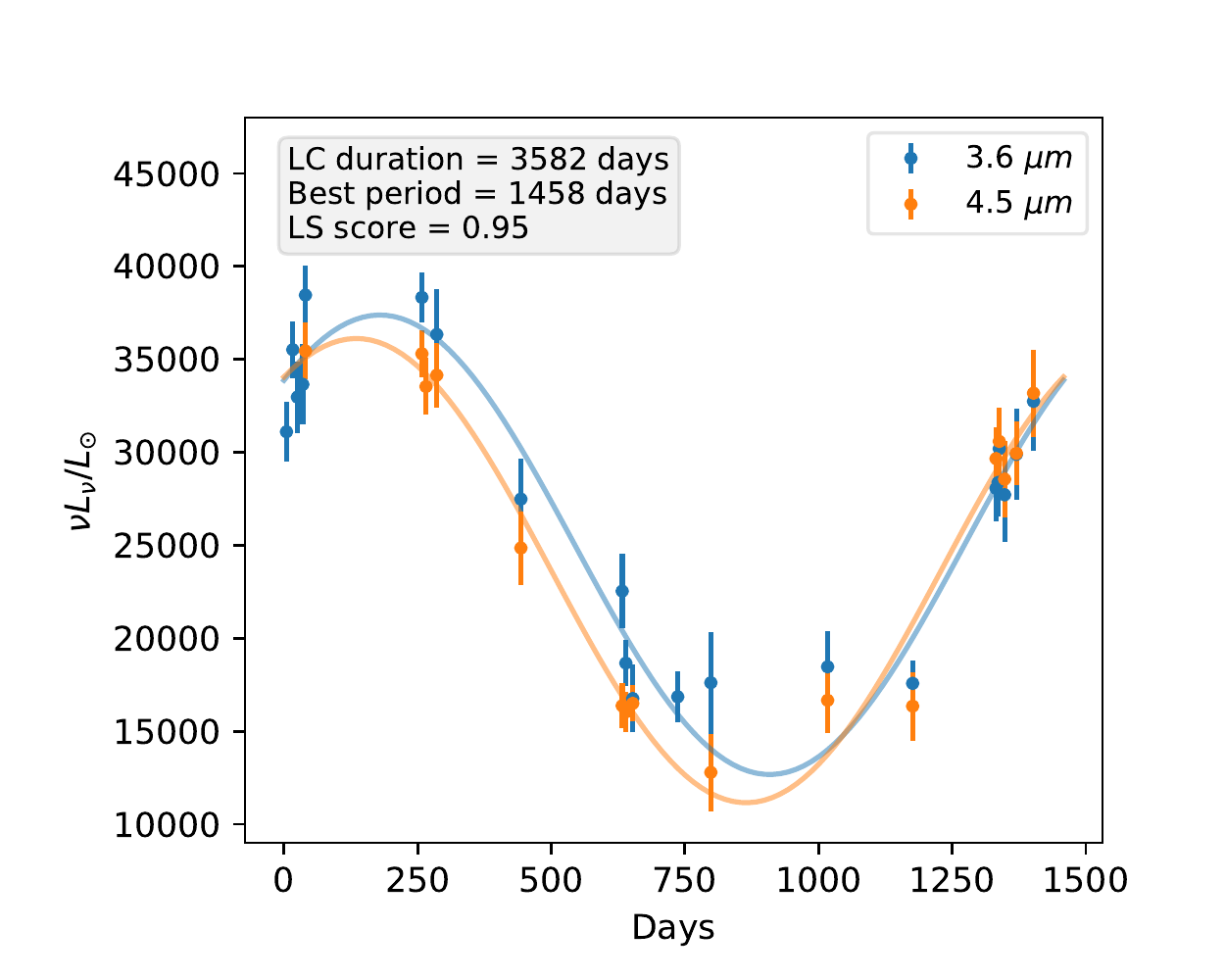}
    \caption{Phase$-$folded light curve of SPIRITS~15kb : NGC~5236. This source belongs to the group of variables around the LMC Masers in Fig.~ \ref{fig:pldiagram} and has $M_{[4.5]} = -12$, $\Delta[3.6]=1.17$ and $\Delta[4.5]=1.28$.}
    \label{fig:15kb}
\end{figure}

The light curves of SPIRITS~14axv (SN2005at Fig. \ref{fig:14axv_lc}) and SPIRITS~14bkv (Fig. \ref{fig:14bkv_lc}) are characterized by the presence of one epoch separated in time distinctly from the remaining light curve, which drives the period finding algorithm. More data are required to identify the period of SPIRITS~14bkv, while SPIRITS~14axv is not a periodic source, but coincident  (within 1.1$\arcsec$) with SN2005at (\citet{Lennarz2012}). Lastly, our data permit reliable determination of periods longer than 250 days only. Hence, it is possible that the primary period in some of our variables is shorter than 250 days and that the value we report is erroneous. The light curve of SPIRITS~14ct (Figs. \ref{fig:14ct_lc} and \ref{fig:14ct}) either has shorter period or erratic variations superimposed on the 2947 days we determine. It also has large amplitude $JHK$ variations  and has been classified as a C-star \citep{Whitelock2013}.  
The source SPIRITS~14aud, for which we determine a period of 694 days, is coincident with the Cepheid V22 having a period of 146 days \citep{Sandage1971}. SPIRITS~14bge is variable and we report a period of 402 days; however, the data (not illustrated) are insufficent for any period to be determined.
These examples indicate the limitations of our classification criteria. 
They are all plotted in Fig.~\ref{fig:pldiagram} at the periods we determined and therefore make a minor contribution to the scatter. 

\clearpage
\newpage


\end{document}